\documentclass[sigconf]{acmart}

\renewcommand\footnotetextcopyrightpermission[1]{} 
\AtBeginDocument{%
  \providecommand\BibTeX{{%
    \normalfont B\kern-0.5em{\scshape i\kern-0.25em b}\kern-0.8em\TeX}}}

\usepackage{tabularx, booktabs, makecell, caption,subfig}
\usepackage{siunitx}
\usepackage[linesnumbered,ruled,vlined]{algorithm2e}
\usepackage{multirow}
\usepackage{colortbl} 
\definecolor{darkcyan}{rgb}{0.0, 0.55, 0.55}

\usepackage{xcolor}

\usepackage{booktabs}

\usepackage{amsmath}
\usepackage{geometry}
\usepackage{xcolor}

\usepackage{cuted}
\usepackage{flushend}

\usepackage{multicol}

\usepackage{subfig}

\usepackage{xcolor}
\usepackage{hyperref}
\hypersetup{colorlinks=true,urlcolor={black!40}}
\newcommand{\deemph}[1]{{\color{black!40}#1}}

\begin{document}

\title{Transferable Graph Backdoor Attack}


\author{Shuiqiao Yang}
\email{shuiqiao.yang@unsw.edu.au}
\affiliation{%
  \institution{The University of New South Wales}
  \streetaddress{P.O. Box 1212}
  \city{Sydney}
  \state{NSW}
  \country{Australia}
  \postcode{}
}

\author{Bao Gia Doan}
\email{bao.doan@adelaide.edu.au}
\affiliation{%
  \institution{The University of Adelaide}
  \streetaddress{}
  \city{Adelaide}
  \state{SA}
  \country{Australia}
  \postcode{}
}

\author{Paul Montague}
\email{paul.montague@defence.gov.au}
\affiliation{%
  \institution{Defence Science and Technology Group}
  \city{Edinburgh}
  \state{SA}
  \country{Australia}
  \postcode{5111}
}

\author{Olivier De Vel}
\email{olivierdevel@yahoo.com.au}
\affiliation{%
  \institution{Defence Science and Technology Group}
  \city{Edinburgh}
  \state{SA}
  \country{Australia}
  \postcode{5111}
}

\author{Tamas Abraham}
\email{tamas.abraham@defence.gov.au}
\affiliation{%
  \institution{Defence Science and Technology Group}
  \city{Edinburgh}
  \state{SA}
  \country{Australia}
  \postcode{5111}
}

\author{Seyit Camtepe}
\email{seyit.camtepe@data61.csiro.au}
\affiliation{%
  \institution{Data61, CSIRO}
  \city{Sydney}
  \state{NSW}
  \country{Australia}
  \postcode{2122}
}

\author{Damith C. Ranasinghe}
\email{damith.ranasinghe@adelaide.edu.au}
\affiliation{%
  \institution{The University of Adelaide}
  \streetaddress{}
  \city{Adelaide}
  \state{SA}
  \country{Australia}
  \postcode{}
}

\author{Salil S. Kanhere}
\email{salil.kanhere@unsw.edu.au}
\affiliation{%
  \institution{The University of New South Wales}
  \streetaddress{P.O. Box 1212}
  \city{Sydney}
  \state{NSW}
  \country{Australia}
  \postcode{}
}


\begin{abstract}

Graph Neural Networks (GNNs) have achieved tremendous success in many graph mining tasks benefitting from the  message passing strategy that fuses the local structure and node features for  better graph representation learning. 
Despite the success of GNNs, and similar to other types of deep neural networks, GNNs are found to be vulnerable to  unnoticeable perturbations on both graph structure and node features. Many adversarial attacks have been proposed to disclose the fragility of GNNs under different perturbation strategies to create adversarial examples. However, vulnerability of GNNs to successful backdoor attacks was only shown recently.

In this paper, we disclose the \textbf{TRAP} attack, a \textbf{T}ransferable G\textbf{RAP}h  backdoor attack.
The core attack principle is to \textit{poison} the training dataset with perturbation-based triggers that can lead to an \textit{effective and transferable backdoor attack}. The perturbation trigger for a graph is generated by performing the perturbation actions on the graph structure via a gradient based score matrix from a surrogate model. Compared with prior works, TRAP attack is different in several ways: i)~it exploits a surrogate Graph Convolutional Network (GCN) model to generate perturbation triggers for a blackbox based backdoor attack; ii)~it generates sample-specific perturbation triggers which do not have a fixed pattern; and iii)~the attack transfers, \textit{for the first time in the context of GNNs}, to different GNN models when trained with the forged poisoned training dataset. Through extensive evaluations on four real-world datasets, we demonstrate the effectiveness of the TRAP attack to build transferable backdoors in four different popular GNNs using four real-world datasets.

\end{abstract}



\maketitle
\pagestyle{plain}


\section{Introduction}

Graphs represents an important data structure to model complex real-world relationships and are used in many domains from social networks  to chemistry, leading to many important applications such as toxic molecule classification, community detection, link prediction and malware detection \cite{yang2021variational}.
Recently, Graph Neural Networks (GNNs) have achieved great success in graph-structured data processing by  learning effective graph representations  via message passing strategies, which recursively aggregate features from neighboring nodes \cite{kipf2016semi,hamilton2017inductive, velivckovic2017graph, xu2018powerful}.
GNNs have outperformed many traditional machine learning techniques for graph processing and become the dominant method for many graph mining tasks \cite{zhou2020graph,yang2021variational}.

On the one hand, GNNs can achieve superb performance compared with other traditional graph mining methods, as they  combine node features and graph structure to learn better representations.
For example, Graph Convolutional Networks (GCN) \cite{kipf2016semi} exploit feature propagation via the adjacency matrix of graphs for  node representation learning.
On the other hand, the intrinsic attribute of the graph-structured data  can be easily attacked as the attacker can easily  manipulate the local structure of a victim node in a graph to infect others ({\em e.g.,}  a fake connection or user profile in a social network) and thus affect the performance of GNNs.
What is more, the message passing strategy of many GNN models can also become an important vulnerability under adversarial attacks \cite{zang2020graph} via graph structure perturbation.
Similar to other type of deep neural networks, GNNs have been shown to be vulnerable and can be easily attacked via unnoticeable perturbation on the graph structure or node features \cite{sun2018adversarial,jin2020graph}.

In recent years, many works \cite{zugner2018adversarial,zugner2019adversarial,wan2021attacking, wu2019adversarial,dai2018adversarial,zang2020graph} have investigated the vulnerability of GNNs to adversarial attacks, where the attackers either modify the graph structure or perturb the node features or both to degrade the performance GNNs on tasks such as graph-level classification, node-level classification, link prediction, etc. 
Depending on different type of tasks (inductive or transductive), such adversarial attacks can happen at   training or inference phase where the adversarial graphs are used to degrade the performance of GNN  models \cite{dai2018adversarial, wang2019attacking, zugner2018adversarial,zugner2019adversarial, zang2020graph}.
For example, Dai et al. \cite{dai2018adversarial} proposed to use reinforcement learning to determine the perturbation actions (adding or deleting edges) for modifying a graph via querying the feedback from a target GNN classifier for both graph-level and node-level classification tasks.
Z{\"u}gner et al. \cite{zugner2018adversarial} proposed Nettack, which exploits incremental
computations to perturb an attributed network with respect to node features and structures and influence the prediction for a target node. 
Later, Z{\"u}gner et al. \cite{zugner2019adversarial} proposed to perturb a graph based on the meta-gradient of the graph {\em w.r.t.} a surrogate model, and then use the poisoned graph to train new GNNs from scratch to degrade their performance on node classification task.
Zang et al. \cite{zang2020graph} proposed Graph Universal Adversarial Attacks (GUA), whose  core idea is to connect a victim node to a bad anchored node and thus misguide the prediction.

\begin{figure}[htb!]
    \centering
    \includegraphics[scale=0.38]{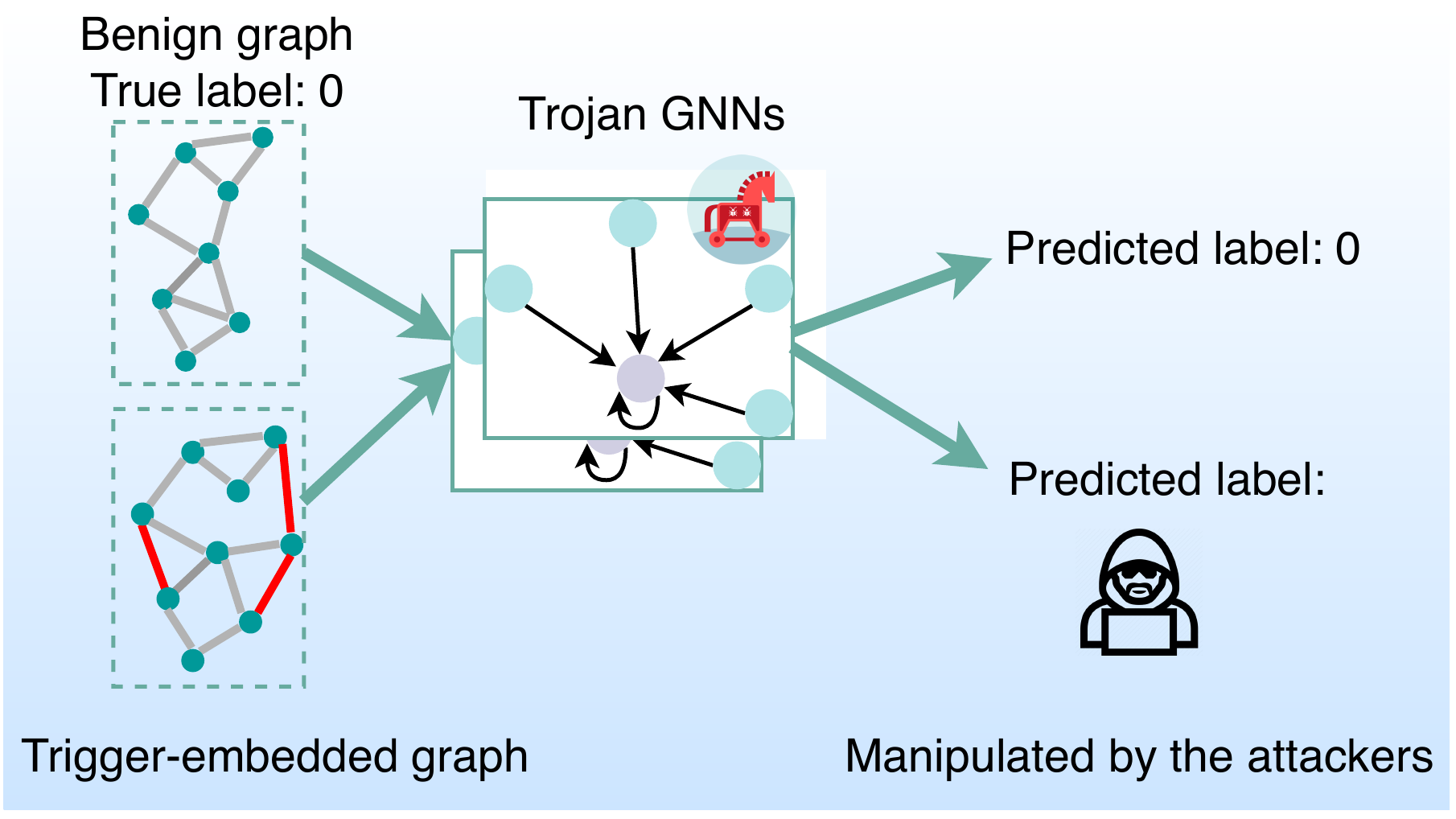}
    \caption{An Illustration of a graph backdoor attack. A trojan GNN model behaves normally on a benign graph but makes an attacker defined prediction on a trigger-embedded graph.}
    \label{fig:graph_backdoor_illustration}
\end{figure}

\begin{figure*}[ht]
    \centering
    \includegraphics[scale=0.3]{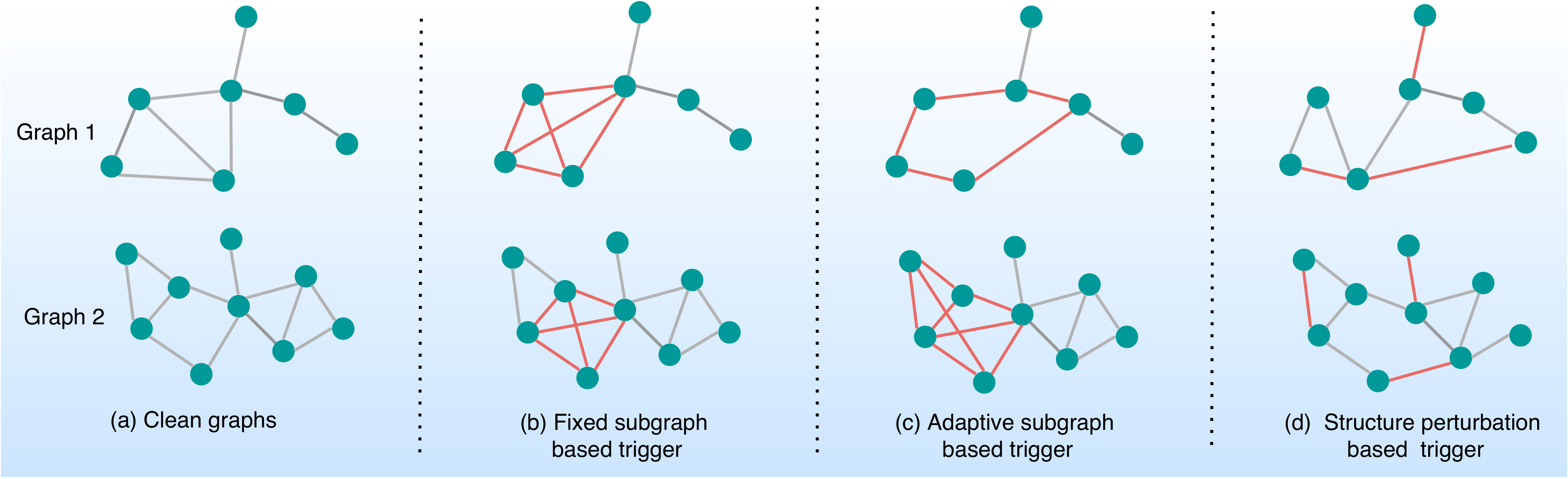}
    \caption{Illustration of trigger-embedded graphs generated by different attacks:  (a) original clean graphs; (b) subgraph  backdoor by Zhang et al. \cite{zhang2021backdoor} adopts the same subgraph as a trigger for all graphs; (c) GTA \cite{xi2021graph} generates subgraphs with different structure for different graphs;  (d) TRAP adopts a trigger generation method via structure perturbation akin to adversarial example generation.}
    \label{fig:trigger_embedded_graphs}
\end{figure*}

Despite the plethora of prior adversarial attacks  proposed for GNNs, the threats of backdoor attacks to GNNs have been rarely explored.
Backdoor attacks intend to inject a maliciously hidden functionality into the deep learning models.
The backdoor model would behave normally on benign inputs, but the hidden backdoor will be activated to mislead the model when the attack-defined trigger is presented \cite{gu2019badnets}. Fig. \ref{fig:graph_backdoor_illustration} illustrates the graph backdoor attack, where a trojan GNN manifests the attacker desired decision on a poisoned graph but performs normally on a clean graph.
Backdoor attacks have been widely explored in computer vision and natural language processing domains.
For example, using a small patch as a trigger could cause the trained image recognition model to malfunction \cite{liu2017neural}.
But only several  backdoor attacks against GNNs have been proposed in recent years \cite{xi2021graph,zhang2021backdoor}. 
For example, Zhang et al. \cite{zhang2021backdoor} proposed a subgraph based backdoor attack against GNN models which exploits  subgraph as universal trigger to poison the training graphs and attacks the graph classification task (We refer this method as \textbf{Subgraph Backdoor} for simplicity).      
Xi et al. \cite{xi2021graph} proposed Graph Trojaning Attack (GTA) which also uses subgraphs as triggers for graph poisoning.
But unlike Subgraph Backdoor \cite{zhang2021backdoor}, GTA learns to generate adaptive subgraph structure for a specific graph.
Different from Subgraph Backdoor and GTA, TRAP learns to generate perturbation trigger, which is adaptive and flexible to different graphs. 
Fig. \ref{fig:trigger_embedded_graphs} illustrates the trigger-embedded graphs based on different backdoor attacks on GNNs.
As we can see, subgraph backdoor attack uses the same subgraph as trigger to poison different graphs.
GTA generates  adaptive subgraphs as triggers for different graphs, where the triggers have the same number of nodes but different structures.
We can observe the following characteristics in the current backdoor attacks on GNNs: 

\begin{enumerate}
    \item In Subgraph Backdoor \cite{zhang2021backdoor}, the attacker uses the same sugbraph structure as a trigger for poisoning different graphs. Hence, the trigger might not be able to optimally poison different graphs in backdoor attack.
    The experimental study in~\cite{zhang2021backdoor} shows that the subgraph trigger needs to be very dense to be informative and effective for injecting a  backdoor into a GNN.
    \item The GTA~\cite{xi2021graph} methods, instead, learns to generate adaptive triggers  for graph poisoning. The attack, whilst being effective, is reliant on white-box access to the victim model (assumes having fully control of the targeted GNN model, {\em i.e.}, model parameters and architecture) to inject the backdoor.
\end{enumerate}



\begin{figure*}[ht]
    \centering
    \includegraphics[scale=0.45]{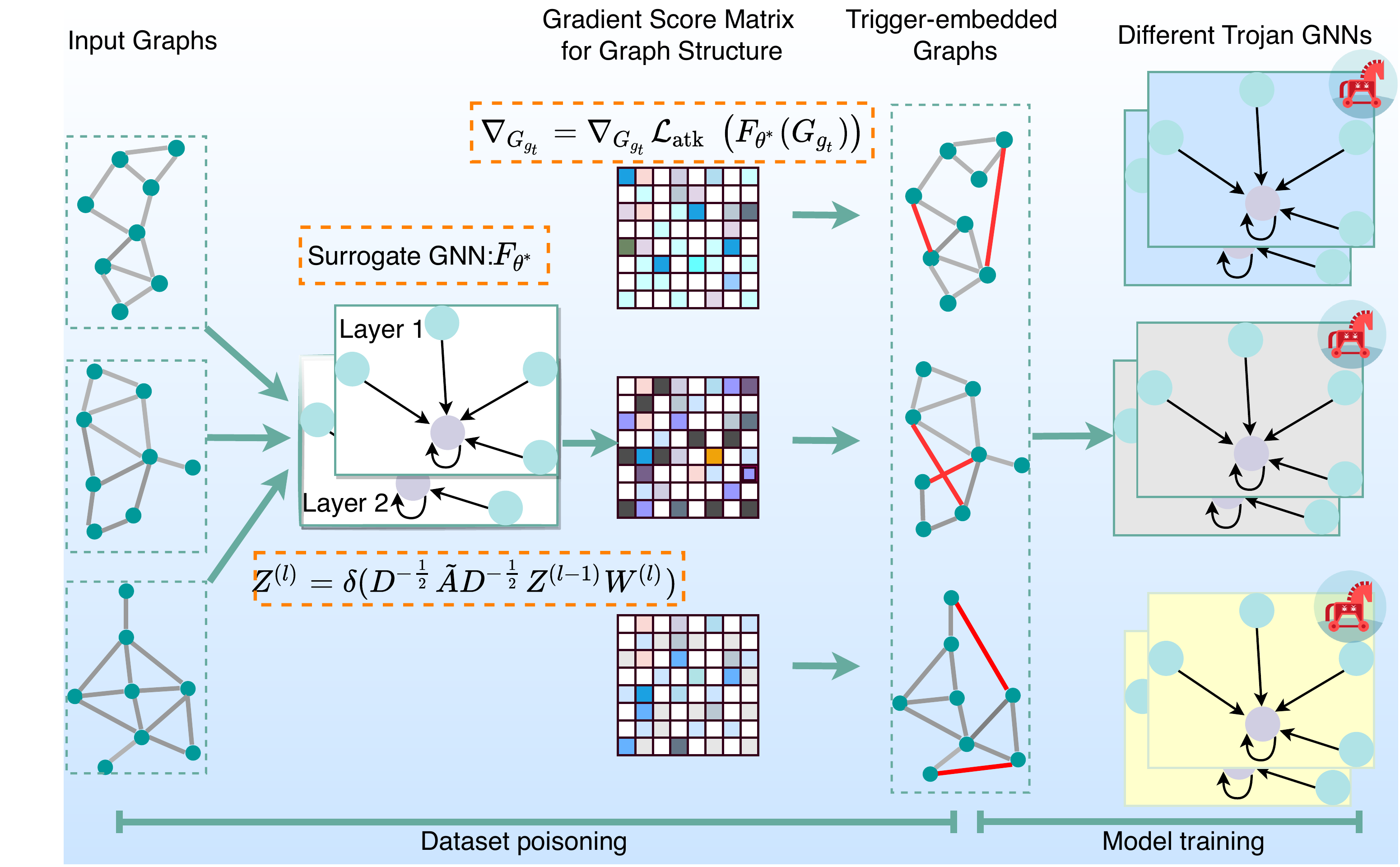}
    \caption{TRAP attack framework. TRAP forges perturbation based triggers for graphs to  poisons the training dataset. A hidden neural Trojan or backdoor will be embedded into a  victim's GNN once the victim adopts the poisoned dataset to train their GNN model.}
    \label{fig:attack_idea}
\end{figure*}

In contrast, we present TRAP attack, a method to generate adaptive perturbation based triggers for a black-box and transferable backdoor attack against GNNs. Compared with Subgraph Backdoor, TRAP learns to generate adaptive perturbation based triggers for different graphs to attain  better attack effectiveness than a universal subgraph trigger. Compared with GTA, TRAP does not require control over the attacked models. On the contrary, TRAP attack exploits a surrogate GNN model to generate trigger-embedded graphs for data poisoning and then relies on the transferability of the trigger to inject a backdoor into a victim's GNN model.
Fig. \ref{fig:attack_idea} shows our attack framework, where the attacker has control over the training dataset, but does not require access to the GNN models adopted by the victims.

The generated trigger is adaptive to different graphs without specific patterns.
To determine the best trigger for a graph, we propose a gradient based score matrix to determine the positions of perturbing the graph structure that can lead to the best attack effectiveness.
The adversarial  trigger-embedded  graphs generated by TRAP turn out to be  transferable and effective in attacking different GNN models such as Graph Convolutional Networks (GCN) \cite{kipf2016semi}, Graph Attention Networks (GAT) \cite{velivckovic2017graph} and Graph Isomorphism Networks (GIN) \cite{xu2018powerful}.
Our contributions are summarized as follows: 
\begin{itemize}
    \item We propose a \textit{new attack} on GNNs; TRAP attack--a graph structure perturbation based backdoor attack on GNNs.
    \item 
    The TRAP attack method exploits a surrogate model to learn to generate perturbation-based triggers that are sample-specific and pattern flexible to poison a small faction of the training data. The poisoned data capable of injecting a backdoor into the surrogate model is shown to successfully transfer to an unseen victim model to enable injecting a backdoor into the model. Hence, TRAP attack can be mounted in a \textit{black-box} setting, and is a \textit{first} demonstration of a transferable backdoor attack in GNNs. This is significant because data-driven model building pipelines of GNNs are reliant on, often, publicly available training data, and now, that data may be poisoned to inject a backdoor into the learned model. 
    \item We extensively evaluate our attack on four real-world graph datasets from various domains and test the transferability of the attack across to four different GNNs. Our empirical results show that the poisoned training dataset generated by the TRAP attack can achieve better attack effectiveness than relevant baselines in regards to transferable backdoor attacks. 
\end{itemize}

\section{Related Work}

Graph Neural Networks (GNNs) have received much attention in recent years and  played  a critical role in many domains.
But similar to the other types of deep  neural networks, GNNs are also vulnerable to malicious attacks. 
Based on the literature, there are two popular types of attacks against GNNs: adversarial attacks and backdoor attacks. 
Here, we briefly review the two different attacks in the following sections.

\subsection{Adversarial Attacks}

Adversarial attacks forge adversarial examples with unnoticeable perturbation on the raw data samples to degrade the performance of the trained models.
To launch adversarial attacks,  attackers require to have knowledge of the targeted model ({\em i.e.,} white-box attack) \cite{sun2018adversarial} or have access to query feedback of the targeted model ({\em i.e.,} black-box attack) to forge adversarial samples  \cite{liu2016delving,papernot2017practical}. 
For example, some adversarial attacks forges  adversarial examples by exploiting the  gradients of the trained deep learning models with respect to their inputs. 
These attacks include FGSM \cite{goodfellow2014explaining}, JSMA \cite{papernot2016limitations}, Deepfool \cite{moosavi2016deepfool} and PGD \cite{madry2017towards}.

Most of the above adversarial attacks have been applied to domains such as computer vision and natural language processing \cite{chakraborty2018adversarial}.
With the development of Graph Neural Networks in recent years, the vulnerabilities of GNNs under adversarial attacks begin to attract researchers' attention.
Some adversarial attacks have been proposed to attack GNNs on different tasks such as link prediction, node-level classification, graph-level classification, graph representation learning, etc.
The graph adversarial examples are forged by perturbing the original graph structure  ({\em i.e.,} adding or deleting an edge), or node features ({\em i.e.,} flipping node feature), or both graph structure and node features \cite{zugner2018adversarial,xu2019topology,wu2019adversarial}.
Existing gradient-based adversarial attacks have been adapted to the graph domain. 
For example, Wu et al. \cite{wu2019adversarial} proposed  integrated gradients based on FGSM and JSMA to overcome the challenges in perturbing the discrete graph structure or features for crafting adversarial examples.
The  integrated gradients are used as guidance to decide the positions for perturbing edges or features.
In addition, other gradient-based attack methods such as Nettack \cite{zugner2018adversarial} have been proposed to forge perturbations on graph structure and node features to create adversarial samples to perform attack on Graph Convolutional  Networks (GCN)  \cite{kipf2016semi}.

\subsection{Backdoor Attacks}

Backdoor attacks attempt to mislead the deep learning models via embedding a hidden malicious functionality ({\em i.e.,} backdoor) into the affected model.
Backdoor attacks appear in both the training phase and inference phase.
During the training phase, a hidden malicious functionality is embedded into the targeted model by poisoning a small ratio of training samples with a special pattern called a trigger.
The trigger can be either sample-agnostic ({\em e.g.,} same  patch or sticker for different images) or sample-specific ({\em i.e.,}  adaptive perturbation noise for different images).
During the inference phase, the backdoor model behaves normally for benign inputs with the absence of a trigger. But as long as the inputs are embedded with a trigger, the backdoor would be activated and mislead the model's prediction.
Backdoor attacks have been mainly studied in the computer vision domain.
For example, Gu et al. \cite{gu2017badnets} first revealed the backdoor threat is realistic in deep neural networks and demonstrated backdoor attacks via using a small sticker as a trigger to mislead an image classifier to label a stop sign as a speed limit sign.
Liu et al. \cite{liu2017neural} also revealed the threats of neural trojans by using a simple patch into a digit number dataset as trigger to train a backdoor model. 
Liao et al. \cite{liao2018backdoor} proposed to use small static or adaptive perturbation  as invisible triggers to attack a CNN model for image classification.
Similarly, Li et al. \cite{li2021invisible} proposed to using auto-encoder to  generate sample-specific invisible noise as trigger via encoding an attacker specified string to poison images.   
Zhao et al. \cite{zhao2020clean} proposed to apply backdoor attacks into video recognition models with 
universal adversarial perturbations and hidden backdoor triggers into clean-label poisoned data samples.
Mu{\~n}oz-Gonz{\'a}lez et al. \cite{munoz2019poisoning} and Bao et al.~\cite{bao2021tnt} exploited GANs to generate triggers, the latter focusing on spatially constrained naturalistic trigger generation while the former investigated generating adversarial training samples.

In addition to adversarial attacks on GNNs, backdoor attacks against GNNs have also been proposed in recent years.
For example, Zhang et al. \cite{zhang2021backdoor} proposed to use subgraph as universal trigger to poison the training dataset and train backdoor GNN models.
Their backdoor attack is targeted, where the labels of the  poisoned graphs are manipulated to the targeted attack label.
Similar to the backdoor attacks in the other domains, the special subgraph trigger pattern is embedded into GNNs as a hidden backdoor. 
The backdoor GNNs would predict a  graph as the desired class as long as the same sugbraph trigger is embedded. 
They tested their  attack transferability on three GNN models ({\em i.e.,} GIN, SAGPool and HGP-SL) and the results showed that the attack  effectiveness heavily relies on the  trigger size.
Xi et al. \cite{xi2021graph} proposed Graph Trojaning Attack (GTA) for targeted backdoor attacks on GNNs. 
Compared with \cite{zhang2021backdoor}, GTA creates adaptive  subgraphs as trigger for different graphs. 
Specifically, GTA iteratively train a trigger generator  with a pre-trained GNN which used to forge backdoor. 
To make sure that GTA is effective on the trigger-embedded graphs and evasive on benign graphs.
They have two constraint when training a trojan GNN:
(1) the  embeddings of benign graphs generated by the trojan GNN are forced to be similar to the embedding generated by the pre-trained clean GNN; (2) the embeddings of poisoned graphs generated by the trojan GNN are  forced to be similar with the embedding of benign graphs in the targeted class.
They formulated the two constraints as a bi-level optimisation problem, and adopted $L_2$ distance as the loss function to train a trojan GNN model.


\section{Problem Formulation}

We consider the task of graph classification, which has been widely applied in security-critical domains such as toxic chemical classification and malware call graph classification.
Given a set of graphs as training data with labels, the goal of graph classification is to train a graph classifier (using Graph Neural Networks) and infer the class of an unlabeled graph.
Formally, let $\mathcal{D}$ be a graph dataset containing $n$ graph  instances $\{(G_1,y_{1}), \dots, (G_n, y_{n})\}$, where $G_i$ is the $i^{th}$ graph and $y_{i}$ is one of the $K$ labels in the label space $\mathcal{Y} = \{{c_1,c_2, \dots, c_K \}}$. 
The goal is to learn a function $F: \mathcal{G} \rightarrow {\mathcal{Y}}$, which maps each unlabeled graph $G$ to one of the K classes in $\mathcal{Y}$.
A graph $G$ contains  two parts: $A$ is the adjacency matrix and $X$ is the node feature matrix.

Note that the mapping function $F$ contains two modules in our setting: $f$ is normally the representation learning module ({\em i.e.,} Graph Neural Networks ) and $h$ is the classifier module ( {\em e.g.,} a fully connected neural network). 
The parameters of $F$ are learned by gradient-descent based optimization with a loss function $\mathcal{L}_{train}$  ( {\em e.g.,} cross-entropy) on the labeled dataset $\mathcal{D}$ as:

\begin{equation}
\theta^{*}=\underset{\theta}{\arg \min } ~ \mathcal{L}_{\text {train }} ((h \circ f)_{\theta}(\mathcal{D})).
\end{equation}

\subsection{Threat Model}
\noindent \textbf{Attacker's goal}. Backdoor attacks inject malicious functions as hidden neural trojans into the target model, and the hidden trojan would be activated and mislead the model to make a desired output when the pre-defined patterns, called triggers, are present.
In our work, the attacker's goal is to exploit backdoor attacks to impact the GNN models trained by end users with two objectives \cite{xi2021graph}: (1) the backdoor GNN models should have normal accuracy on benign graphs; (2) the backdoor GNN models should make desired decisions on trigger-embedded graphs.

Suppose we have a backdoor GNN model $F= (h \circ f)_{\theta}$ and a clean GNN model $F_{\circ} = (h \circ f)_{\theta_\circ}$. 
The two objectives of the attack can be formally defined as follows:

\begin{equation}\label{eq: attack_obj}
\left\{\begin{array}{l}
 F (G) = F_{\circ}(G) \\  
F(G_{g_t})=y_{t}, 
\end{array}\right.
\end{equation}
where $y_t$ is the targeted attack class and ${g_t}$ denotes a trigger. $G$ and $G_{g_t}$ represents a clean graph and a trigger embedded graph, respectively.
From Eq. \ref{eq: attack_obj} we can see that the first objective specifies the evasiveness of a trojan GNN model, {\em i.e.,}  that the  trojan and clean models behave similarly on the clean graphs, 
while the second objective represents the attack effectiveness, {\em i.e.,}  that the trojan model will predict the trigger-embedded graph in the targeted attack class.
\vspace{2mm}

\noindent 
\textbf{Attacker's knowledge}. To launch backdoor attacks, the attacker can have different levels of knowledge about the target, such as access to the training data $\mathcal{D}$, the parameters and structure of the target GNN models $F_{\theta}$ used by users, the classifier $h$, etc. 
Usually, if the attacker requires less knowledge to the target, then the attacks are also more realistic and dangerous.
In our work,  we focus on  more realistic attacks and assume the attacker has limited knowledge as shown in the following: 
\begin{itemize}
    \item The architecture and parameters of the GNN models are agnostic to the attacker. 
    \item The attacker could poison a small ratio of the training dataset with triggers.
\end{itemize}

Our attack aims at poisoning the training data with trigger-embedded graphs for backdoor attacks, and do not assume any knowledge on the targeted GNN models.

\vspace{2mm}
\noindent 
\textbf{Attacker's capability}.
As the attacker has no access to the targeted GNN models, we thus use a surrogate model as the object of our attack.
We use the surrogate model to tune the trigger generation and poison the training dataset.  
After our attack is effective on the surrogate model, we then publish the poisoned training dataset to transfer the backdoor attack on GNN models that will be adopted by the end users. 
In order to achieve the attack goals ({\em i.e.,} effectiveness and evasiveness), our attacks have to meet the two requirements: 
\begin{itemize}
    \item The poisoned graphs should be transferable to different unseen GNN models. 
    \item The triggers embedded in the poisoned graphs should be unnoticeable to users. 
\end{itemize}

To meet the first requirement, our attack forges a trigger to poison a graph by perturbing the graph structure at the edge positions where the message passing of a surrogate GNN model would be maximally influenced.
To meet the second requirement, different from \cite{xi2021graph} and \cite{zhang2021backdoor} where a trigger is a subgraph, our attack generates an adaptive trigger for a graph which does not have specific pattern and has small perturbation size.

\subsection{Surrogate GNN model}

Graph Neural Networks (GNNs) have demonstrated great expressiveness in  graph representation learning by fusing both graph structure and node features with the message passing strategies over the graph edges. 
Moreover, GNNs can also implicitly aggregate features from $k$-hop neighbors with multi-layers. 
There are many variants of GNNs. We adopt a Graph Convolutional  Networks (GCN) \cite{kipf2016semi} as our surrogate model to launch the backdoor attack, considering the universality of its message passing strategy.
Specifically,  for a graph $G = \{A, X\}$  where $A$ is the adjacency matrix and $X$ is the node attribute matrix, the graph representation learning of a multi-layer GCN is performed as follows:

\begin{equation}\label{eq:gcn_layer}
    Z^{(l)} = \delta(D^{-\frac{1}{2}} \tilde{A} D^{-\frac{1}{2}} Z^{(l-1)} W^{(l)}),
\end{equation}
where $\tilde{A}  = A + I$, $D$ is the degree matrix based on $\tilde{A}$, $\delta$ is an activation function such as $ReLU(\cdot) = max(0, \cdot)$, and $W^{(l)}$ is the trainable parameters for the $l^{th}$ GCN layer.
$Z^{(l)}$ is the leaned representations for the graph nodes. 
As shown in Eq. \ref{eq:gcn_layer}, GCN aggregates the node features from the local neighborhood structure through the normalized adjacency matrix.
Since GCN is original designed for node classification, it learns the representation of a graph at the node level. In order to classify a graph, we use a pooling layer ({\em e.g.,} max pooling)  to get the graph-level representation and then use a simple fully connected neural network layer with softmax for classification.

\section{Attack Method}

In this section, we detail our proposed TRAP attack.
From a high-level aspect, the core idea of TRAP is to construct trigger-embedded graphs to poison the training dataset.
Once the poisoned dataset is used to train a new GNN model from scratch, the trained model will be embedded with a backdoor.



\subsection{Attack Design}

Suppose we have a clean graph dataset $\mathcal{D}$ which can be partitioned into two parts: $\mathcal{D}[y_t]$ and $\mathcal{D}[\neg y_t]$. 
The candidate graphs for embedding trigger  are randomly sampled from $\mathcal{D}[\neg y_t]$, whose true labels are not the targeted attack label $y_t$.
To forge trigger-embedded graphs, our attack goals can be formulated as the following  optimization problem:

\begin{equation}\label{eq: opti_task}
\begin{array}{c} 
\underset{{G}_{g_t} \in \Phi(G)}{\min}  \mathcal{L}_{\text {atk}}\left(F_{\theta^{*}}(G_{g_t})\right)\quad 
\text{s.t.} \quad \theta^{*}=\underset{\theta}{\arg \min} ~   \mathcal{L}_{\text {train }}\left(
F_{\theta}(\mathcal{D})\right), 
\end{array}
\end{equation}
where $\mathcal{L}_{atk}$ is the loss function used by the attacker for constructing poison graphs while $\mathcal{L}_{train}$  is the loss function for training a graph classifier (surrogate model in our attack). 
$\Phi(G)$ denotes an admissible  space for constructing a trigger-embedded graph with a given constraint.
Recall that the attacker can manipulate the label of the poisoned graphs and replace them with the targeted attack class. 
Also, the ratio of poisoned graphs compared with the clean training graphs is low ({\em e.g.,} 5\%).
Once we have a trained model, $F_{\theta^*}$, a candidate graph $G$ for poisoning tend to be classified by $F_{\theta^*}$ as its true label.
Hence, the attacker needs to generate a trigger on the graph that can mislead the $F_{\theta^*}$ under a limited budget.
Therefore, the loss function of $L_{atk}$ can be same as $L_{train}$ as the training label of a poisoning $G$ for attack is already manipulated as $y_t$.

Unlike the previous graph backdoor attacks \cite{xi2021graph,zhang2021backdoor} that regard the trigger as a subgraph,  the trigger in our attack is more flexible and without a specific shape.
We generate an adaptive trigger for a graph in  adversarial style by perturbing the structure of the graph via adding or deleting edges that lead to the maximum decrease of the attacking loss $\mathcal{L}_{atk}$.
Meanwhile, we limit the edge perturbation size so that the changes on poisoned graphs are unnoticeable.
The benefit of perturbing the graph structure to forge trigger are two-fold:
(1) Since GNNs rely on message passing on the graph structure for feature aggregation, perturbing the structure may lead to transferable triggers.
(2) A perturbation based graph trigger at flexible edge positions shows randomness and does not have specific patterns, which could be stealthier.

To determine the best edge position to perturb, an intuitive way is to  treat the graph adjacency matrix as  hyper-parameter and compute its gradients  with respect to the  attack loss:


\begin{equation}\label{eq:gradient_gragh_structure}
\begin{array}{c}
\nabla_{G_{g_t}} = \nabla_{G_{g_t}} \mathcal{L}_{\text {atk }}\left(F_{\theta^{*}}(G_{g_t})\right) \\ 
\quad \text{ s.t. } \quad \theta^{*}=\underset{\theta}{\arg \min} ~  \mathcal{L}_{{train}}\left(F_{\theta}{(D)})\right) 
\end{array}
\end{equation}

The best model parameter $\theta^*$ is normally determined after training the surrogate model via stochastic gradient decent as: 

\begin{equation}
\theta_{t+1}=\theta_{t}-\alpha [\nabla_{\theta_{t}} \mathcal{L}_{\text{train }}\left(F_{\theta_{t}}(D)\right)]
\end{equation}

Inspired by previous works that use meta-gradient for adversarial attacks \cite{zugner2019adversarial,yuan2021meta} which  indicate that interactions between the hyperparameters and model parameters are important, the gradient for $G_{g_t}$ in our attack can also consider the impact of parameters $\theta^*$  which is computed as: 

\begin{equation}
\begin{array}{l}
\nabla_{G_{g_t}} = \nabla_{F} \mathcal{L}_{\text {atk }}\left(F_{\theta^{*}}(G_{g_t})\right)[\nabla_{G_{g_t}}F_{\theta^{*}}(G_{g_t}) + \nabla_{\theta^*}F_{\theta^*}(G_{g_t})\nabla_{G_{g_t}}\theta^*] \\\\ 
\end{array}
\end{equation}

We then  update $G_{g_t}$ based on $\nabla_{G_{g_t}}$. 
A straightforward way is to perturb $G_{g_t}$  with gradient descent:  $G_{g_t}'  = G_{g_t} - \beta \nabla_{G_{g_t}}$.
However, unlike the continuous domains ({\em e.g.,} images), graphs are unstructured and discrete, so directly applying the gradient-based update is not suitable for graphs.
Moreover, due to the limited budget for perturbing a graph, we also can not perform a global update on the graph structure. 
Thus, we propose to partially perturb the graph structure based on the gradient with a simple strategy to maintain the graph discreteness and maximize the attack performance under a limited budget.

\subsection{Perturb Graph Structure  for  Trigger Embedding}

Recall that we treat the graph structure as a hyper-parameter and compute the gradient of the attack loss with respect to it. Thus, for a graph $G$ with node size $n$, the shape of the gradient matrix is $\nabla_{G_{g_t}} \in \mathcal{R}^{n\times n}$, where each entry denotes the gradient between two nodes.
Considering the discreteness of the graph, the perturbation on a graph structure can only have two possible cases: Delete an existing edge or add a new edge between two nodes.
Similar to the adversarial attack in  \cite{zugner2019adversarial}, we define a score function to combine the gradient and the adjacency matrix, and  assign the possible perturbation between each node pair a numerical value to indicate the change.
Specifically, for a given node pair $(u,v)$ with $a_{uv}$ as its existing edge status ({\em i.e.,} 0 or 1) in the graph adjacency matrix. 
The score function is defined as $\mathcal{S}(u,v) = \nabla_{G_{g_t}}^{uv} \cdot (2 a_{uv} -1)$.  
The score function retains the sign of the node pair's gradient if there is an existing edge such that its edge can be changed in the gradient direction or vice versa. 
Then, we greedily pick the $M$ highest scores in $\mathcal{S}$ for perturbation, where $M$ is our attack budget ({\em i.e.,} number of edge perturbations). 
The updates meet the requirement of the discreteness of graph structure via adding or deleting an edge given the current status of $a_{uv}$. 
Algorithm \ref{algorithm} sketches our attack flow.




\begin{algorithm}[!t]
\SetAlgoLined
\KwIn{Graph dataset $\mathcal{D}$, targeted attack class $y_t$, attack budget $M$, training epochs $T$, learning rate $\alpha$; 
}

\KwOut{Poisoned graph set $\{G_{g_t}\}$.}

${G_{g_t}}$ $\leftarrow$ randomly sample candidate graphs from $\mathcal{D}[\neg y_t]$; 

Change label of graphs in $\{G_{g_t}\}$ to $y_t$; 

Initialize  surrogate model $F_{\theta_0}$; 

\For{$t$ $\leftarrow$ 0 \KwTo $T-1$ }
{
$\theta_{t+1}$ $\leftarrow$  $\theta_{t}$ - $\alpha$ $\nabla_{\theta_{t}}$ $\mathcal{L}_{train}(\mathcal{D})$;
}

\For{$G_{g_t}$ $\in$ $\{G_{g_t}\}$}
{

\deemph{// Eq. \ref{eq:gradient_gragh_structure}, calculate the gradient of graph structure} 
$\nabla_{G_{g_t}} = \nabla_{G_{g_t}} \mathcal{L}_{\text {atk}}\left(F_{\theta^{*}}(G_{g_t})\right)$;

\deemph{// Gradient based score matrix for graph structure}

$\mathcal{S} \leftarrow \nabla_{G_{g_t}} \cdot (2  G_{g_t} - 1)$;

\deemph{// Find  $M$ number of entries for perturbation}

$\{(u,v)\}_1^M \leftarrow$ get $M$ number of maximum node pairs based on score $S$; 

Add or delete edges $\{(u,v)\}_1^M$ for $G_{g_t}$;
}

\deemph{// The trigger-embedded graphs will be used to poison training dataset}

\Return trigger-embedded graphs $\{G_{g_t}\}$. 
\caption{TRAP Attack}
\label{algorithm}
\end{algorithm}


\section{Attack Evaluation}

In this section, we conduct  empirical studies to answer the following research questions: 
\begin{itemize}
\item[]\textbf{Q1}: Is TRAP effective on graph classification task? 
\item[]\textbf{Q2}: Is TRAP transferable to different GNNs?
\item[]\textbf{Q3}: Is the transferability of TRAP affected by  the structure of the attacked GNN?
\item[]\textbf{Q4}: What is the impact of data poisoning rate on  TRAP? 
\item[]\textbf{Q5}: What is the impact of edge perturbation size to  TRAP?
\end{itemize}



\subsection*{Experimental Settings}

\begin{table*}[h]
    \centering
    \caption{Dataset statistics.}
    \begin{tabular}{c|c | c | c | c |c |c} \hline 
    \toprule
        Dataset & \# Graphs & \# Nodes (Avg.)  &  \#  Edges (Avg.) & \# Classes & \# Graphs in Class & Target Class     \\ 
        \midrule 
        FRANKENSTEIN & 4,337 & 16.90 & 17.88 & 2 &  1936 [0], 2401 [1] & 0\\ \hline 
        Fingerprint  & 1,459 & 8.92 & 7.55 & 3 & 472 [0], 536 [1], 451 [2]& 2\\ \hline
        PROTEINS & 1,113 & 39.06 & 72.82 & 2 & 663 [0], 450 [1] & 1\\ \hline
        WinMal  & 1,361 & 781.87 &1849.41 & 2 & 546 [0], 815 [1] & 0\\ \hline

    \end{tabular}
    \label{tab:datasets}
\end{table*}

\noindent \textbf{Datasets.~}We evaluate TRAP with four real world datasets.  
Table \ref{tab:datasets} shows the basic statistics for each dataset, including the graph numbers in the dataset, the average number of nodes per graph, the average number of edges per graph, the number of classes, the number of graphs in each class and the target class for attack.
We choose the class which has least data samples as the target class for attack.

FRANKESTEIN, Fingerprint and PROTEINS are collected from TUDATASET \cite{morris2020tudataset}, which is a collection of graph datasets for graph classification and regression.
FRANKENSTEIN is created by the fusion of the BURSI and MNIST datasets, where BURSI corresponds to 4337 molecules, with 2401 mutagens
and 1936 nonmutagens \cite{orsini2015graph}. 
Fingerprint is a collection of fingerprints formatted as graph structures from the NIST-4 database \cite{watson1992nist}. 
PROTEINS is a set of macromolecules used to predict  the existence of  enzymes in a protein molecule \cite{dobson2003distinguishing}.
WinMal is a collection of  API call graphs extracted from 1,361 Windows PE (portable executable) files, including 815 malware and 546 goodware \cite{ranveer2015comparative}.

\vspace{2mm}
\noindent \textbf{Metrics. ~} We evaluate the compared methods with two metrics to evaluate the effectiveness and evasiveness of the attacks. 
The first metric is Attack Success Rate (ASR), which measures the success rate with which a backdoor GNN model predicts a trigger-embedded graph into the designated class:

\begin{equation}
    \text{Attack Success Rate (ASR)} = \frac{\# ~ \text{successful attacks}}{\#~\text{total trials}}. 
\end{equation}

The second metric is Clean Accuracy Drop (CAD), which measures the accuracy difference between a clean GNN model and a trojan GNN model in prediction on clean data samples. 

\vspace{2mm}
\noindent \textbf{Baselines.~} We compare TRAP attack with two state-of-the-art baselines. 

\begin{itemize}
    \item Subgraph backdoor by Zhang et al. \cite{zhang2021backdoor} embeds a predefined  subgraph as a universal trigger into training graphs to backdoor GNN models. The subgraph trigger is controlled by the graph size, the  density and  synthesis method.
    \item Graph Trojaning Attack (GTA) by Xi et al. \cite{xi2021graph} trains a trigger generator to  generate  adaptive subgraphs as a trigger for a graph. 
    The trigger generator is trained  together with a pre-trained target GNN model by iteratively updating parameters. 
    The training process will continue until the trigger generator generates triggers that can successfully backdoor the pre-trained GNN model.
\end{itemize}

\vspace{2mm}
\noindent \textbf{Attacked GNN models.~}
We investigate the  transferability of  TRAP to different GNN models and assume that the attacked GNN models are agnostic. 
We choose four popular GNN models: Graph  Convolutional Networks (GCN) \cite{kipf2016semi}, Graph Isomorphism Network (GIN) \cite{xu2018powerful}, GraphSAGE (GSAGE) \cite{hamilton2017inductive} and Graph Attention networks (GAT)\cite{velivckovic2017graph} as the attack models.
Table \ref{tab:clean_model_acc} shows the clean model accuracy on different datasets and GNN models.

\begin{table}[h]
\centering
    \caption{Clean model accuracy.}
\begin{tabular}{c|ccccl}\toprule
Dataset & \multicolumn{4}{c}{GNN models} \\\cmidrule(lr){2-5}
           & GCN & GIN  & GSAGE & GAT \\\midrule
FRANKENSTEIN & 0.6737 & 0.6194 & 0.5899 & 0.6275   \\
Fingerprint  & 0.8235 & 0.8356 & 0.6947 & 0.8767 \\
PROTEINS  & 0.7195 & 0.7014 & 0.6995 & 0.7354 \\
WinMal   & 0.8322 & 0.8220 & 0.8066 & 0.8791\\
\bottomrule
\end{tabular}
\label{tab:clean_model_acc}
\end{table}

To launch attacks, we poison the training dataset with trigger-embedded graphs. Then, the poisoned training dataset will be used to train and backdoor the GNN models. 
Subgraph  backdoor \cite{zhang2021backdoor} generates a subgraph as a universal trigger to poison the training dataset. 
For the GTA \cite{xi2021graph} and TRAP, a surrogate GNN model is required to forge an adaptive trigger into a graph. 
We choose GCN as the surrogate model for both GTA and TRAP.
By default, we apply the same structure settings (\textit{i.e.}, the number of layers  and neurons) between surrogate model and victim GNN models to test the attack transferability.

\vspace{2mm}
\noindent \textbf{Dataset splits and construction.~} The same data split rules are applied to all the methods.
We randomly split the dataset into three parts: 70\% is used for a clean training dataset, 20\% is used as a clean testing dataset and the remaining 10\% will be used to embed triggers (poisoned graphs).

The graphs used for trigger embedding are chosen from the non-targeted classes, and their label will be manipulated into the target class.
Half of the poisoned graphs will be mixed with the clean training dataset to construct a poisoned training dataset, and the remaining poisoned graphs will be used for testing at inference time.

\vspace{2mm}
\noindent \textbf{Parameter settings.~} All the adopted GNN models contain a two-layer (the neuron sizes are 16 and 8 in each layer) structure followed by a  max pooling layer for node level feature aggregation, and a fully connected layer with softmax for graph classification.
The subgraph trigger size for  and GTA is set as 5 nodes. 
Subgraph Backdoor adopts the Erdős-Rényi (ER) \cite{gilbert1959random} model to generate a subgraph as trigger with its trigger density set as 0.8.
GTA adopts a three layer fully connected neural network as trigger generator.  
Adam optimizer with learning rate of 0.01 is used to train  the GNN models.
Table \ref{tbl:parameter_settings}  shows the parameter settings for all the adopted GNNs and our experimental studies.

\begin{table}[h]
\centering
\caption{Parameter settings.}
\begin{tabular}{c| c| c }\toprule
Type  & Parameter  & Settings  \\ \hline 
& & \\ 
 GIN& Architecture & two layers (16, 8) \\
GCN, GSAGE,    & Classifier & FCN + Softmax \\
 GAT & Aggregator & Max Pooling\\ 
& & \\ \hline
GAT   & Number of heads & 3 \\
\hline \hline 
 & Trigger size & 5 \\ 
Subgraph Backdoor & Trigger density & 0.8\\  
 & Trigger generation & ER \cite{gilbert1959random} \\ 

\hline 
& Trigger size & 5 \\ 
GTA & Trigger generator & 3 layers FCN \\ 
\hline 

TRAP & Edge perturbation size & 5 (insertion or deletion) \\
\hline
\hline 

& Optimizer  & Adam\\  
& Learning rate  & 0.02\\ 
Training & Weight decay &  $5e^{-4}$\\
& Batch size & 100 \\ 
 & Epochs & 50 \\ 
& Poisoned  data rate & 5\%  \\ 
 \hline

\bottomrule
\end{tabular}
\label{tbl:parameter_settings}
\end{table}



\begin{table*}[hbt!]
\centering
\caption{Comparison of attack effectiveness on the GCN model (the model employed as the surrogate).  Highest Attack Success Rate (ASR)  and lowest Clean Accuracy Drop (CAD) scores are highlighted.}
\begin{tabular}{|l|c|c|c|c|c|c|} 
\hline
\multicolumn{1}{|l|}{\multirow{2}{*}{Dataset}} & \multicolumn{2}{c|}{Subgraph Backdoor }                & \multicolumn{2}{c|}{GTA}                            & \multicolumn{2}{c|}{TRAP}                        \\ 
\cline{2-7}
\multicolumn{1}{|l|}{}                         & \multicolumn{1}{c|}{ASR} & \multicolumn{1}{c|}{CAD} & \multicolumn{1}{c|}{ASR} & \multicolumn{1}{c|}{CAD} & \multicolumn{1}{c|}{ASR} & \multicolumn{1}{c|}{CAD}  \\
\hline 
FRANKENSTEIN                & 0.8360 & 0.0271    &    0.9429 & 0.0274     & {\cellcolor[rgb]{0.753,0.753,0.753}} 0.9595 &  {\cellcolor[rgb]{0.753,0.753,0.753}} 0.0197                           \\
Fingerprint     & 0.2955 & 0.0147 & 0.6818 & {\cellcolor[rgb]{0.753,0.753,0.753}}  0.0014 & {\cellcolor[rgb]{0.753,0.753,0.753}} 0.7241 & 0.0171   \\
PROTEINS    & 0.3333 & {\cellcolor[rgb]{0.753,0.753,0.753}} 0.0004 & 0.4137 & 0.0084 & {\cellcolor[rgb]{0.753,0.753,0.753}} 0.7778 & 0.0224 \\
WinMal      &0.4909 & {\cellcolor[rgb]{0.753,0.753,0.753}} 0.0150 & 0.1176 & 0.0215 &{\cellcolor[rgb]{0.753,0.753,0.753}} 0.7593 & 0.0183 \\ \hline
\end{tabular}\label{tbl: gcn_results}
\end{table*}

\subsection*{\textbf{Q1}: Is TRAP Effective on Graph Classification Tasks?}

Table \ref{tbl: gcn_results} summarizes the performance of different attacks on the GCN model. 
Overall, TRAP achieves much better attack effectiveness and evasiveness than its counterpart methods. 

Specifically, all the attacks achieve high attack effectiveness on the FRANKENSTEIN dataset. The best attack is from the proposed TRAP, with over 0.95 ASR achieved, while the ASR of Subgraph Backdoor and GTA are 0.8360 and 0.9429, respectively. 
On the other three datasets, TRAP still achieves the highest ASR compared with the other two baselines.
For instance, on Fingerprint, the ASR of TRAP on GCN is 0.7241, which is higher than GTA (0.6818) and Subgraph Backdoor (0.2955).
Similar trends can be found on the PROTEINS and WinMal datasets. The ASR achieved by GTA and Subgraph Backdoor has also dropped significantly.
For example, on the WinMal dataset, the ASR of GTA  on GCN can only achieve 0.1176,  
while TRAP still shows much better ASR (0.7593).

As for the CAD shown in Table \ref{tbl: gcn_results}, we can see that TRAP has the lowest CAD on the FRANKENSTEIN dataset (0.0197), lower than the CAD achieved by Subgraph Backdoor and GTA (0.0271 and 0.0274 respectively).
The CAD of TRAP is less than 2.5\%.
Even though Subgraph Backdoor and GTA show lower CAD on the other thee datasets, they can not achieve better ASR indicating that the attacks from Subgraph Backdoor and GTA are less effective.
Overall, TRAP shows better balance between effectiveness (ASR) and evasiveness (CAD) than Subgraph Backdoor and GTA. 

\subsection*{\textbf{Q2}: Is TRAP Transferable to Different GNNs?}

\begin{table*}[hbt!]
\centering
\caption{Attack Success Rate (effectiveness) comparison. Highest ASR  are highlighted. The structure of surrogate GNN is same with the attacked GNNs. }
\begin{tabular}{|c|c|c|c|c|c|c|c|c|c|} 
\hline 
\multirow{2}{*}{Dataset} & \multicolumn{3}{c|}{Subgraph Backdoor}                   & \multicolumn{3}{c|}{GTA} & \multicolumn{3}{c|}{TRAP}  \\ 
\cline{2-10} 
                         & GAT & GIN & \textcolor[rgb]{0.125,0.129,0.141}{GSAGE} & GAT & GIN & GSAGE        & GAT & GIN & GSAGE              \\ 
\hline

    FRANKENSTEIN &  0.5079 & 0.9206 & 0.8677 & {\cellcolor[rgb]{0.753,0.753,0.753}}  0.9018 & 0.6800 &  0.9523 & 0.8208 & {\cellcolor[rgb]{0.753,0.753,0.753}}0.9823 & {\cellcolor[rgb]{0.753,0.753,0.753}} 1.0\\
Fingerprint  & 0.3596 & 0.3864 & 0.4205 &  0.6939 & 0.4318 & {\cellcolor[rgb]{0.753,0.753,0.753}} 0.6591 &  {\cellcolor[rgb]{0.753,0.753,0.753}} 0.8621 & {\cellcolor[rgb]{0.753,0.753,0.753}} 0.7586 &   0.5862 \\
PROTEINS  & 0.2576 &  0.4697 & 0.3333 & 0.5294 & 0.6552 & 0.5517 & {\cellcolor[rgb]{0.753,0.753,0.753}} 0.5556 &  {\cellcolor[rgb]{0.753,0.753,0.753}} 0.6889 & {\cellcolor[rgb]{0.753,0.753,0.753}} 0.5556\\
WinMal   &  0.40 &  0.60 & 0.5818 & 0.4615 & 0.3056 & 0.1613 & {\cellcolor[rgb]{0.753,0.753,0.753}} 0.7222 & {\cellcolor[rgb]{0.753,0.753,0.753}} 0.6111 & {\cellcolor[rgb]{0.753,0.753,0.753}}  0.6111 \\

\hline
\end{tabular}\label{tbl: attack_success_rate}
\end{table*}

\begin{table*}[hbt!]
\centering
\caption{Clean Accuracy Drop (evasiveness) comparison. Lowest CAD are highlighted. The structure of surrogate GNN  is same with the attacked GNNs.}
\begin{tabular}{|c|c|c|c|c|c|c|c|c|c|} 
\hline
\multirow{2}{*}{Dataset} & \multicolumn{3}{c|}{Subgraph Backdoor}                   & \multicolumn{3}{c|}{GTA} & \multicolumn{3}{c|}{TRAP}  \\ 
\cline{2-10}
                         & GAT & GIN & \textcolor[rgb]{0.125,0.129,0.141}{GSAGE} & GAT & GIN & GSAGE        & GAT & GIN & GSAGE              \\ 
\hline
FRANKENSTEIN    &   0.0360 & 0.0660 &  0.0368 & 0.0192 & {\cellcolor[rgb]{0.753,0.753,0.753}} 0.0155 & 0.0264   & {\cellcolor[rgb]{0.753,0.753,0.753}} -0.0116 &  0.0185 & {\cellcolor[rgb]{0.753,0.753,0.753}} 0.0080 \\
Fingerprint     &  {\cellcolor[rgb]{0.753,0.753,0.753}} -0.0190 & 0.0314 &  0.0047 &  0.1327 & 0.0186 & 0.0413   &  0.024 &{\cellcolor[rgb]{0.753,0.753,0.753}}  {\cellcolor[rgb]{0.753,0.753,0.753}} 0.0 &  {\cellcolor[rgb]{0.753,0.753,0.753}} -0.0136 \\
PROTEINS        &  0.0194 & {\cellcolor[rgb]{0.753,0.753,0.753}} 0.0149 & 0.0407 &  0.0401 & 0.1786 &  0.0862 &{\cellcolor[rgb]{0.753,0.753,0.753}}  0.0134 &  0.0224 & {\cellcolor[rgb]{0.753,0.753,0.753}} 0.009 \\
WinMal         &{\cellcolor[rgb]{0.753,0.753,0.753}}   -0.0229 &  0.0884 & 0.0361  & 0.0116 & 0.0280 & 0.0932 &   0.0036 & {\cellcolor[rgb]{0.753,0.753,0.753}} 0.0183 &  {\cellcolor[rgb]{0.753,0.753,0.753}} 0.0147\\
\hline
\end{tabular}\label{tbl: clean_accuracy_drop}
\end{table*}

In this set of experiments, we investigate the attack transferability of TRAP to different GNN models.
The attacked models have the same structure setting as the surrogate model ({\em i.e.,} two layers with each neuron sizes are set as  16 and 8).

Table \ref{tbl: attack_success_rate} and Table \ref{tbl: clean_accuracy_drop} summarize the attack efficacy of transferability on GAT, GIN and GSAGE  with regard to the  attack effectiveness (ASR) and attack evasiveness (CAD).
The results show that TRAP can achieve much better attack transferability to different GNN models.
However, the backdoor attacks from the two baselines cannot be transferred effectively to different GNN models.

First, we can see from Table \ref{tbl: attack_success_rate} that TRAP achieves better ASR than the other two baselines on most of the datasets. 
For example, on the FRANKENSTEIN dataset, the ASR achieved by TRAP is 0.8208, 0.9823 and  1.0 for GAT, GIN and GSAGE, respectively.
GTA achieves the  best ASR on GAT  and second best on GSAGE, which are 0.9018 and 0.9523, but lowest ASR on GIN which is only 0.68. 
Subgraph Backdoor achieves the second best ASR on GIN, which is 0.9206,  but lower ASR on GAT and GSAGE, which are 0.5079 and 0.8677, respectively.

On the other three datasets, TRAP still achieves the highest ASR compared with the other two baselines, except for GSAGE on Fingerprint. 
For instance, on Fingerprint, the ASR of TRAP method on GAT and GIN are 0.8621 and 0.7586, which are higher than GTA (0.6939 and 0.4318) and Subgraph Backdoor (0.3596 and 0.3864). 
Similar trends can be found on the PROTEINS and WinMal datasets. The ASR achieved by GTA and Subgraph Backdoor has dropped significantly. 
For example, on the WinMal dataset, the ASR of GTA  on GAT, GIN and GSAGE can only achieve 0.4615, 0.3056 and 0.1613, respectively, 
while TRAP still shows much better ASR across the three GNN models ({\em i.e.,} 0.7222, 0.6111 and 0.6111) on the Winmal dataset.

Table \ref{tbl: clean_accuracy_drop} shows the attack evasiveness in regard to clean accuracy drop.
The TRAP method shows overall lower clean accuracy drop compared with  Subgraph Backdoor and GTA, which indicates that the TRAP attack method has better evasiveness. 
On the FRANKENSTEIN dataset, the clean accuracy drop of the TRAP attack is less than 2\% across  different GNN models.
The average CAD of Subgraph Backdoor on FRANKENSTEIN across the GNNs is over 3\% and the highest CAD is over 6\%.
GTA shows slightly lower CAD than Subgraph Backdoor on FRANKENSTEIN. 
For example, the CAD on GIN is 1.5\%, which is the lowest among the three attack methods, but higher CAD on GAT and GSAGE than TRAP.
For some cases, the CAD of GTA  and Subgraph Backdoor is lower than TRAP.
For example, the CAD of trojan GCN and GIN from the Subgraph Backdoor attack on the PROTEINS dataset is lower than the CAD of the TRAP. 
But in most of the cases, TRAP achieves the better balance between attack effectiveness and evasiveness.
Even though the baselines like Subgraph Backdoor can have lower CAD in some cases, their ASR is also lower, we can argue that their backdoor attack is ineffective.


\begin{table*}[hbt!]
\centering
\caption{Attack Success Rate (effectiveness) comparison. Highest ASR  are highlighted. The structure of surrogate GNN is different with the attacked GNNs.}
\begin{tabular}{|c|c|c|c|c|c|c|c|c|c|c|c|c|} 
\hline 
\multirow{2}{*}{Dataset} & \multicolumn{4}{c|}{Subgraph Backdoor}                   & \multicolumn{4}{c|}{GTA} & \multicolumn{4}{c|}{TRAP}  \\ 
\cline{2-13} 
                         &GCN & GAT & GIN & GSAGE & GCN & GAT & GIN & GSAGE       &GCN & GAT & GIN & GSAGE              \\ 
\hline

FRANKENSTEIN &   0.7302 & 0.4497 & 0.8783 & 0.6667 & {\cellcolor[rgb]{0.753,0.753,0.753}} 0.9247 & {\cellcolor[rgb]{0.753,0.753,0.753}} 0.9135 & 0.5534& 0.7272 & 0.8786 & 0.8844 & {\cellcolor[rgb]{0.753,0.753,0.753}} 0.9827 &  {\cellcolor[rgb]{0.753,0.753,0.753}} 0.9769 \\
Fingerprint  & 0.2247 & 0.4270 & 0.2022 & 0.3933 &  {\cellcolor[rgb]{0.753,0.753,0.753}} 0.6957& 0.5909 & 0.3259 &  0.4468  &  0.6379 & {\cellcolor[rgb]{0.753,0.753,0.753}} 0.8448  & {\cellcolor[rgb]{0.753,0.753,0.753}} 0.6034  & {\cellcolor[rgb]{0.753,0.753,0.753}} 0.5517 \\
PROTEINS  & 0.2879 & 0.2424 & 0.1818 & 0.3940 & 0.6207 & 0.1936 &  0.1724 & 0.3125 & {\cellcolor[rgb]{0.753,0.753,0.753}} 0.8 & {\cellcolor[rgb]{0.753,0.753,0.753}} 0.6222  & {\cellcolor[rgb]{0.753,0.753,0.753}} 0.5333  & {\cellcolor[rgb]{0.753,0.753,0.753}} 0.4444\\
WinMal   &  0.3455 & 0.2545 & 0.4  & 0.4 & 0.2121 &  0.1026  & 0.1290 &  0.2941 & {\cellcolor[rgb]{0.753,0.753,0.753}} 0.7778 & {\cellcolor[rgb]{0.753,0.753,0.753}} 0.8889 &{\cellcolor[rgb]{0.753,0.753,0.753}}  0.5370 & {\cellcolor[rgb]{0.753,0.753,0.753}} 0.7037 \\

\hline
\end{tabular}\label{tbl:asr_structure}
\end{table*}

\begin{table*}[hbt!]
\centering
\caption{Clean Accuracy Drop (evasiveness) comparison. Lowest CAD are highlighted. The structure of surrogate  is different with the  attacked GNNs.}
\begin{tabular}{|c|c|c|c|c|c|c|c|c|c|c|c|c|} 
\hline 
\multirow{2}{*}{Dataset} & \multicolumn{4}{c|}{Subgraph Backdoor}                   & \multicolumn{4}{c|}{GTA} & \multicolumn{4}{c|}{TRAP}  \\ 
\cline{2-13} 
                         &GCN & GAT & GIN & GSAGE & GCN & GAT & GIN & GSAGE       &GCN & GAT & GIN & GSAGE              \\ 
\hline

FRANKENSTEIN &   0.027 & 0.0225 &  0.07 &  0.02 &  0.0296 &   0.0524 &  {\cellcolor[rgb]{0.753,0.753,0.753}} -0.0035 &  0.0184  & {\cellcolor[rgb]{0.753,0.753,0.753}} 0.0011 &  {\cellcolor[rgb]{0.753,0.753,0.753}} -0.0185 &  0.0150 &  {\cellcolor[rgb]{0.753,0.753,0.753}} -0.0023 \\
Fingerprint  &  0.0253 &  0.0063 &  {\cellcolor[rgb]{0.753,0.753,0.753}} 0.0127 &  0.0 &  {\cellcolor[rgb]{0.753,0.753,0.753}}  -0.0099 &  {\cellcolor[rgb]{0.753,0.753,0.753}} -0.0005 &  0.0406 &   0.0368 & 0.0170 & 0.0034 & 0.0442 & {\cellcolor[rgb]{0.753,0.753,0.753}} -0.0274 \\
PROTEINS  &  {\cellcolor[rgb]{0.753,0.753,0.753}} 0.0129 &  0.0064  &  0.0194 &  0.0129 &  0.0411 &  0.0768 & 0.024 & 0.0420 & 0.0179 & {\cellcolor[rgb]{0.753,0.753,0.753}} 0.0044 &  {\cellcolor[rgb]{0.753,0.753,0.753}} -0.0090 & {\cellcolor[rgb]{0.753,0.753,0.753}} 0.0090 \\
WinMal   &  0.0367 & 0.0184 &  {\cellcolor[rgb]{0.753,0.753,0.753}} 0.0091 & 0.0459 & 0.0158 &  0.1013  & 0.0307 &  0.0317 &  {\cellcolor[rgb]{0.753,0.753,0.753}} - 0.0037 & {\cellcolor[rgb]{0.753,0.753,0.753}} -0.0183 & 0.0219 &  {\cellcolor[rgb]{0.753,0.753,0.753}} 0.0073\\

\hline
\end{tabular}\label{tbl:cad_structure}
\end{table*}

\subsection*{\textbf{Q3}: Is the Transferability of TRAP Affected by the Structure of the Attacked GNNs?}

In this set of experiments, we evaluate the transferability of the backdoor attack when the attacked GNNs have different structures compared to the surrogate model.
The structure of the surrogate model is set as two layers with neuron sizes set as 16 and 8, and the structure of the attacked GNN models is set as two layers with neuron sizes set  32 and 16.

Tables \ref{tbl:asr_structure} and \ref{tbl:cad_structure} show the attack success rate and clean accuracy drop, respectively. 
For the attack effectiveness, we can see in Table \ref{tbl:asr_structure} that our TRAP attack still outperforms the other two baselines in  attack success rate in most of the cases.
GTA  has higher ASR than the TRAP attack, but only on the datasets FRANKENSTEIN and PROTEINS when attacking GCN and GAT.
The Subgraph Backdoor performs the worst among the three attacks.
Specifically, on FRANKENSTEIN, TRAP shows good overall ASR across the four different GNNs. The ASR scores are 0.8786, 0.8844, 0.9827 and 0.9769 on GCN, GAT, GIN and GSAGE, respectively. 
Compared with the TRAP, GTA reaches better ASR scores on FRANKENSTEIN with GCN and GAT, which are 0.9247 and 0.9135, but lower ASR scores with GIN and GSAGE, which are 0.5534 and 0.7272.
On Fingerprint, the ASR scores of TRAP are 0.6379, 0.8448, 0.6034 and 0.5517 with GCN, GAT, GIN and GSAGE. 
GTA shows higher ASR (0.6957) on Fingerprint with GCN but lower ASR with the
other GNN models.
Both Subgraph Backdoor and GTA show decreased ASR scores when the attacked GNNs have increased the complexity (bigger neuron sizes). 
Subgraph Backdoor use a universal trigger to poison the graphs  and does not require a surrogate model to generate triggers. 
As the GNN model increases in complexity, the universal trigger based attack becomes weak. 
On the other hand, when the attacked GNNs have different structures to the surrogate model, the attack from GTA becomes less effective. This indicates that the backdoor attack from GTA is sensitive to the GNN structure. 

As for the clean accuracy drop shown in Table \ref{tbl:cad_structure}, we can see that the TRAP has lower drop in most of the cases. Even though the GTA and Subgraph Backdoor attacks show lower CAD scores in some cases, TRAP shows better balance between the two objectives: effectiveness and evasiveness.

\subsection*{Q4: What is the Impact of Data Poisoning Rate on TRAP?}
\begin{figure*}[hbt!]
    \centering
    \subfloat[FRANKENSTEIN\label{<FRANKENSTEIN>}]{\includegraphics[width=0.25\textwidth]{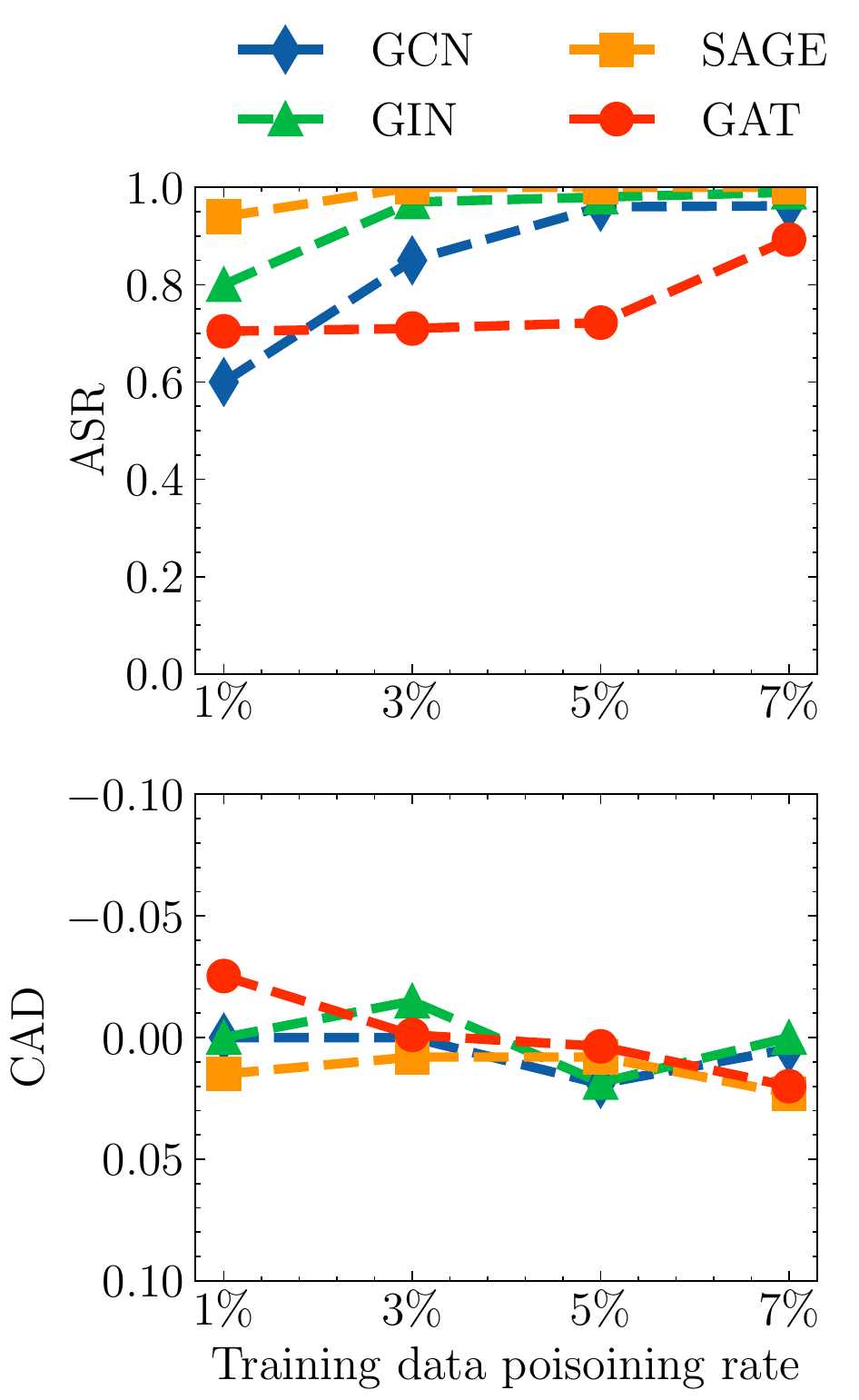}}\hfill
    \subfloat[Fingerprint\label{<Fingerprint>}]{\includegraphics[width=0.25\textwidth]{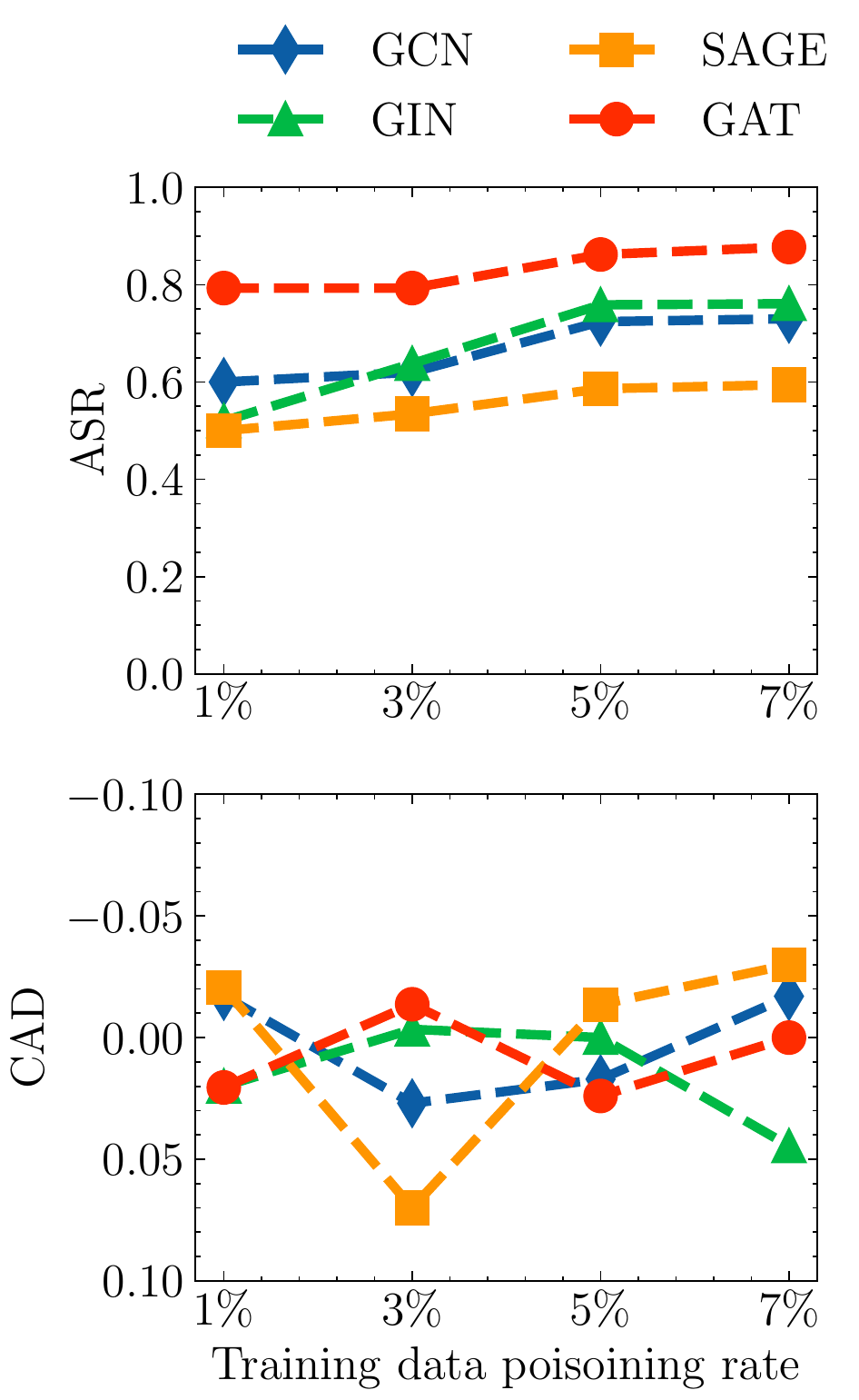}}\hfill    
    \subfloat[PROTEINS\label{<PROTEINS>}]{\includegraphics[width=0.25\textwidth]{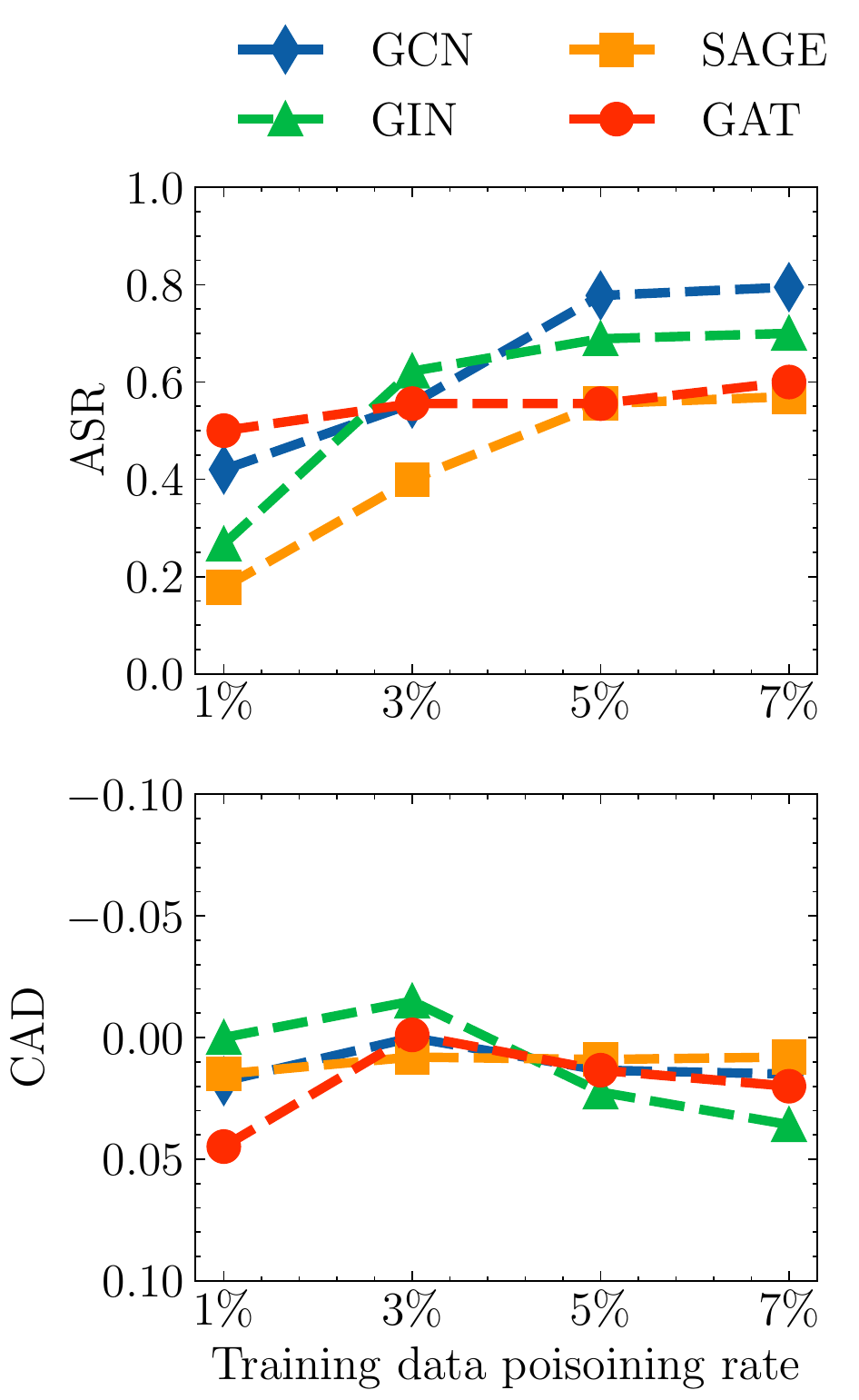}}\hfill    
    \subfloat[WinMal\label{<WinMal>}]{\includegraphics[width=0.25\textwidth]{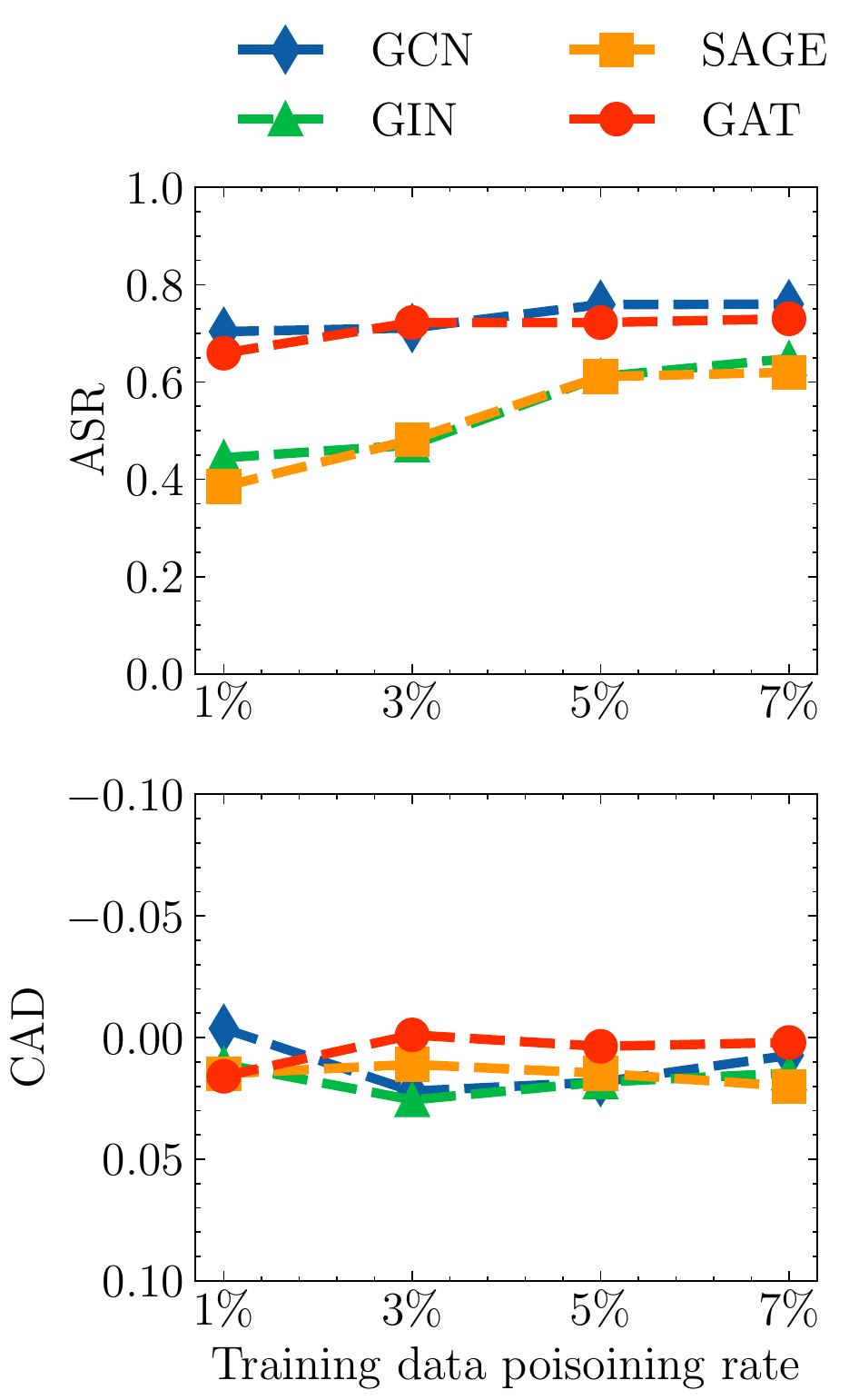}}\hfill  
     \caption{Impact of training data poisoning rate on attack effectiveness and evasiveness.}
     \label{fig:poisoning_rate}
\end{figure*}

This set of experiments evaluate the impacts of data poisoning rate on the effectiveness and evasiveness of TRAP.
Intuitively, more trigger-embedded graphs contained in the  training data would make the  neural networks become better in recognizing the  pattern of the trigger.
Fig. \ref{fig:poisoning_rate} shows the attack success rate and clean accuracy drop based on different training data poisoning rates on the adopted datasets. 
The ASR of TRAP monotonically increases with the training data poisoning rate, and the trend is consistent on different datasets.
Take FRANKENSTEIN as an example. When the poisoning rate is 1\%, the ASR on GCN, GSAGE, GIN and GAT is around 0.60, 0.80, 0.95 and 0.70, respectively.
When the data poisoning rate increases to 7\%, we can see that the ASR on GCN, GSAGE, GIN and GAT increases to 0.96, 1.0, 1.0 and 0.89, respectively.
The ASR shows a sharp increase when the poisoning rate increased from 1\% to 5\%. 
This trend can  be found on other datasets such as FRANKENSTEIN and PROTEINS.
But the ASR does not increase much when the poisoning rate is increased from 5\% to 7\%. 
This phenomenon indicates that TRAP does not rely on higher data poisoning rate to be successful. On the contrary, the generated adversarial style trigger pattern is quite poisonous for effectively training a backdoor GNN model. 

As for the influence of poisoning rate on clean accuracy drop, we can see that the CAD does not show a consistent pattern with the increasing of poisoning rate. In most of the cases,  the clean accuracy drop is less then 2\%. 
A higher data poisoning rate does lead to a big CAD. On the contrary, the CAD keeps at a similar level on most datasets when the data poisoning rate changes from 3\% to 7\%.

\subsection*{Q5: What is the Impact of Edge Perturbation Size on TRAP?}

\begin{figure*}[hbt!]
    \centering
    \subfloat[FRANKENSTEIN\label{<FRANKENSTEIN1>}]{\includegraphics[width=0.25\textwidth]{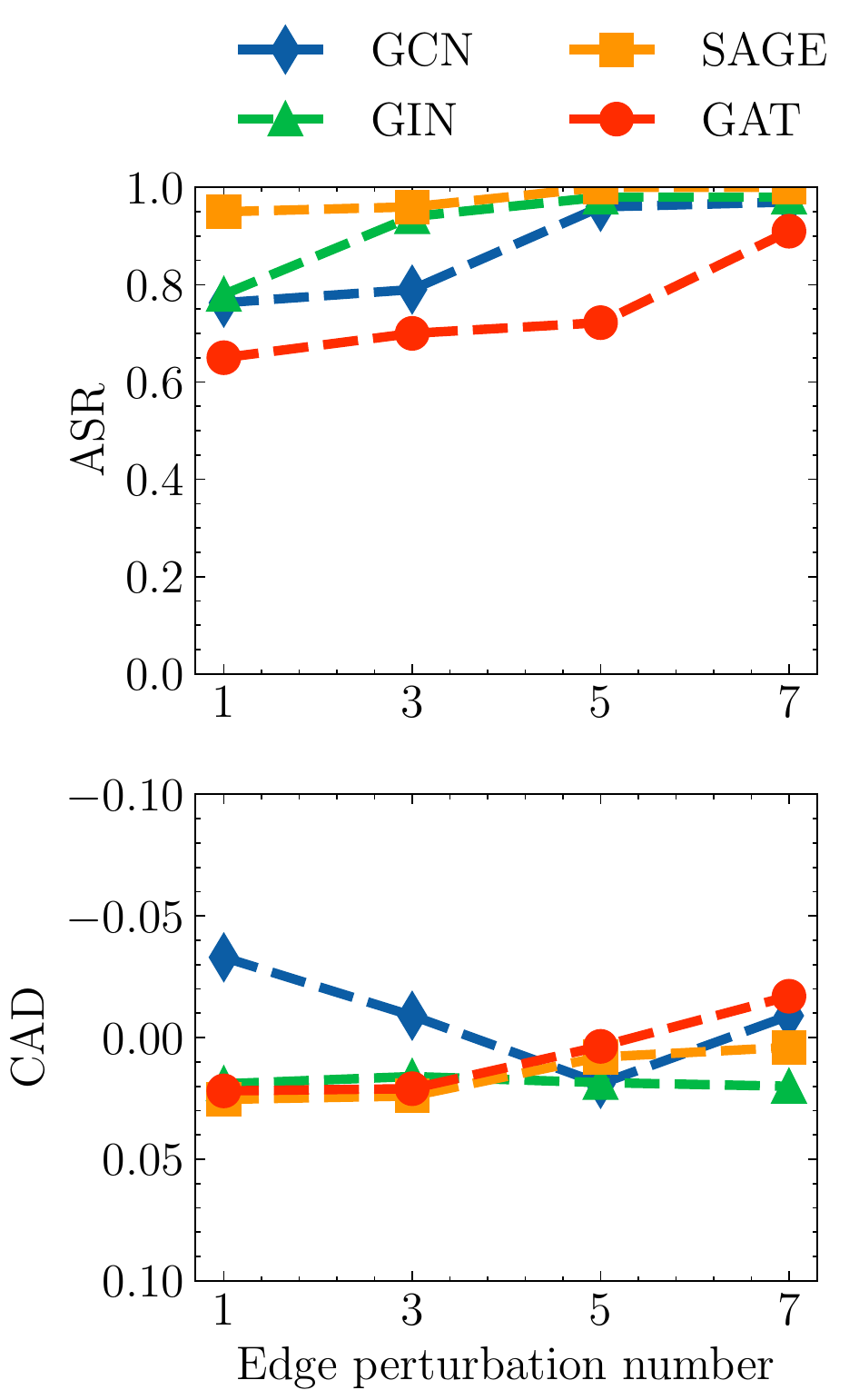}}\hfill
    \subfloat[Fingerprint\label{<Fingerprint1>}]{\includegraphics[width=0.25\textwidth]{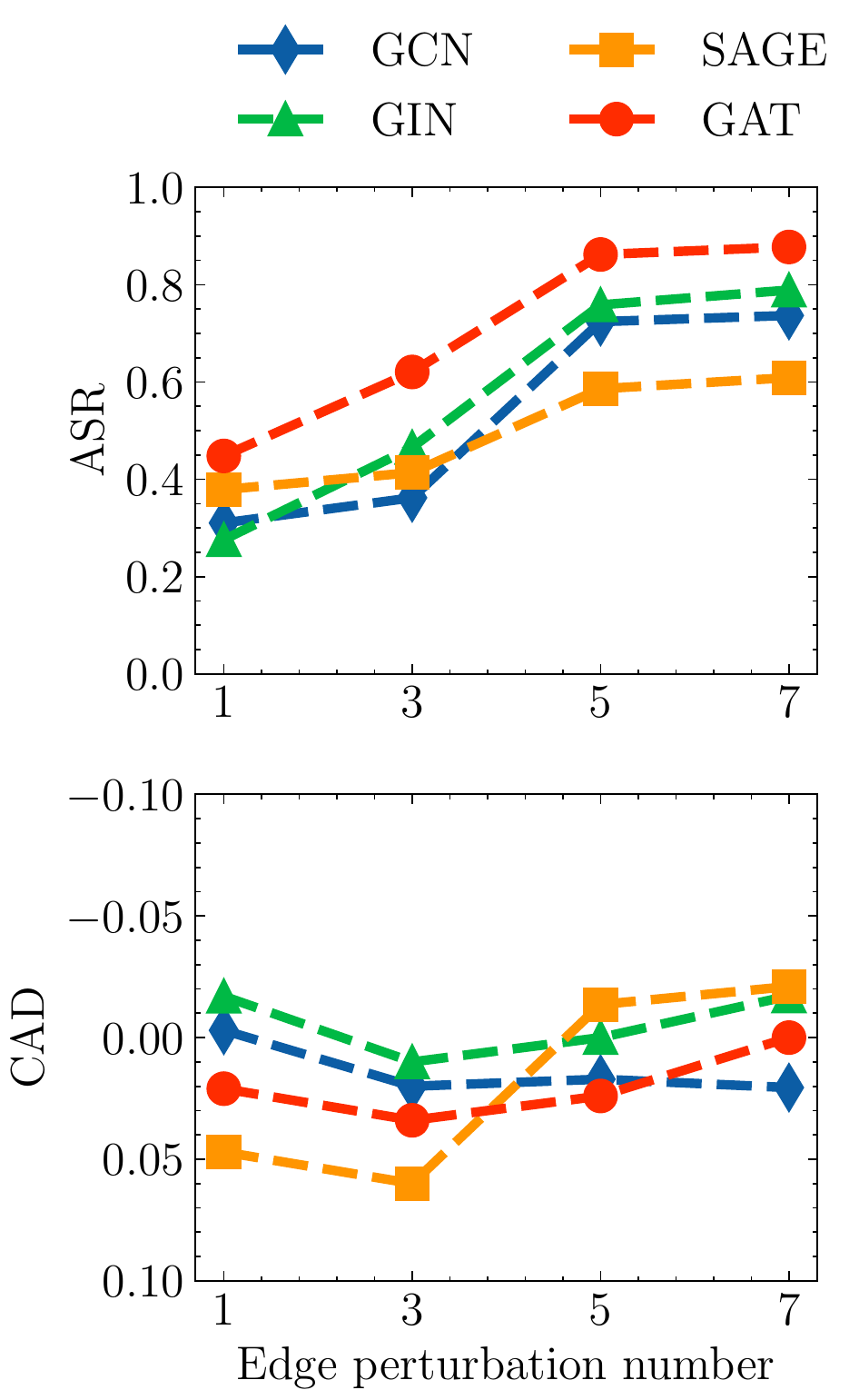}}\hfill    
    \subfloat[PROTEINS\label{<PROTEINS1>}]{\includegraphics[width=0.25\textwidth]{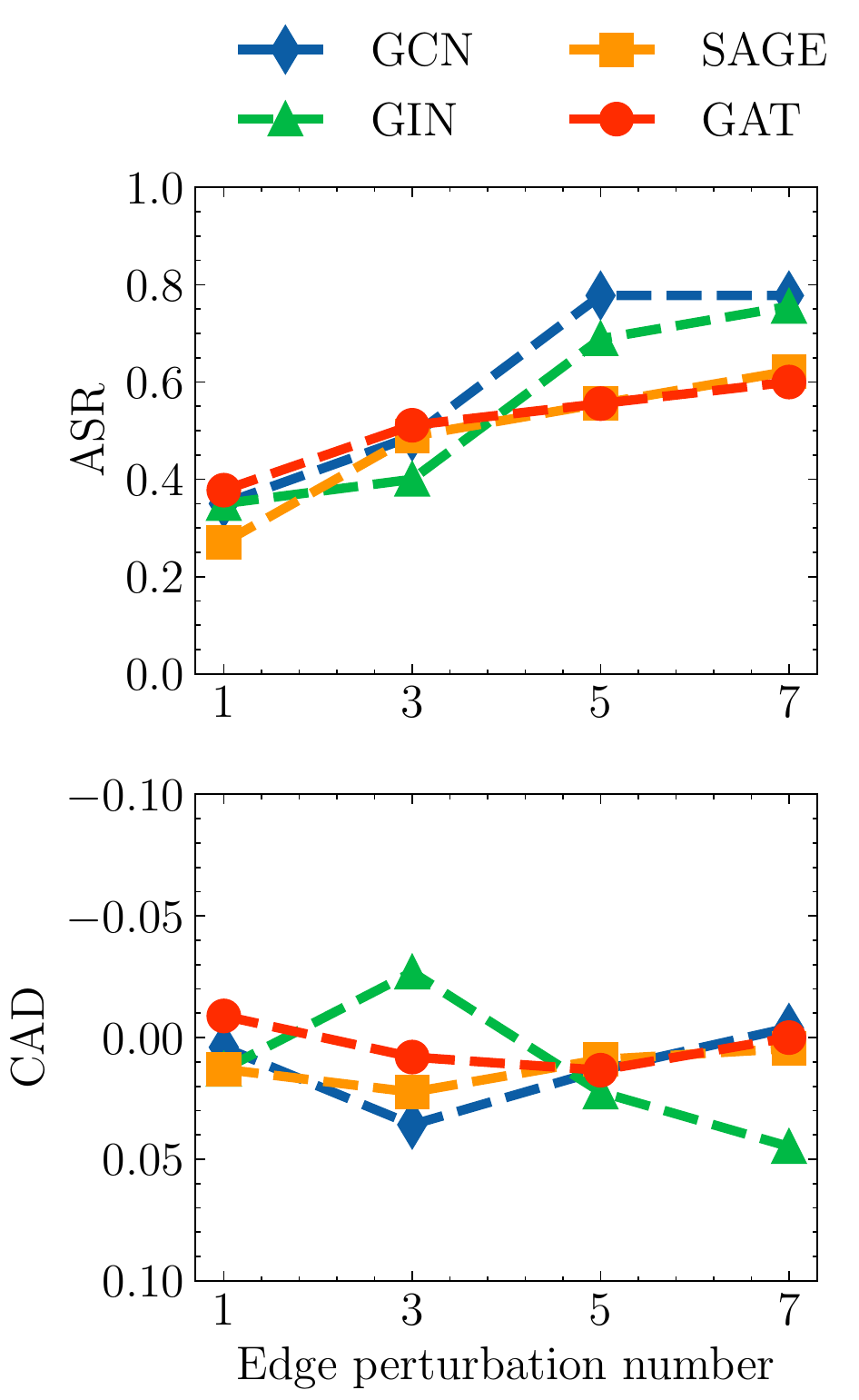}}\hfill    
    \subfloat[WinMal\label{<WinMal1>}]{\includegraphics[width=0.25\textwidth]{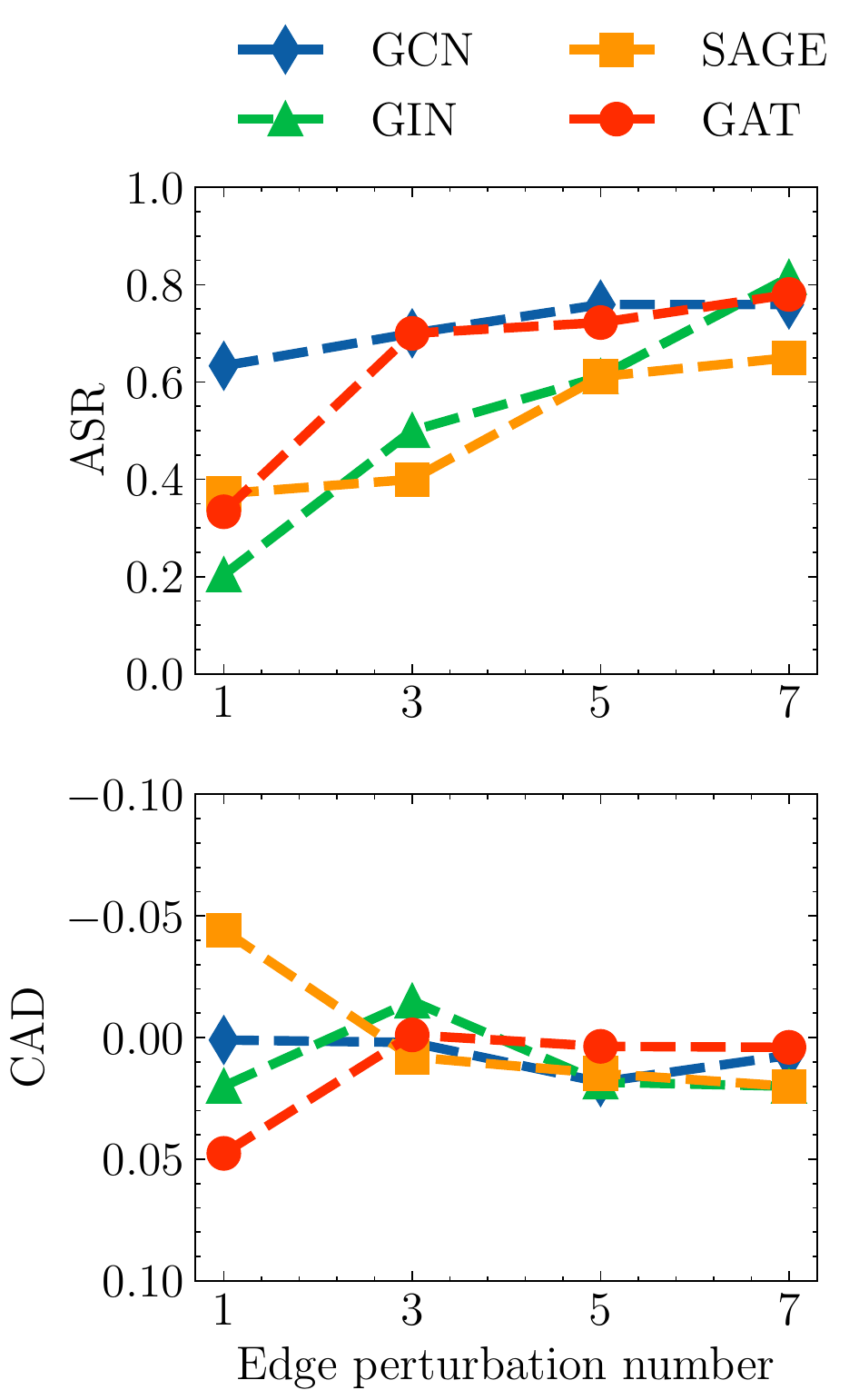}}\hfill  
     \caption{Impact of edge perturbation number on attack effectiveness and evasiveness.}
     \label{fig:trigger}
\end{figure*}

This set of experiments evaluate the impacts of  edge perturbation size  on the effectiveness and evasiveness of TRAP.
Fig. \ref{fig:trigger} shows the impact of edge perturbation size on the attack effectiveness and evasiveness.
Normally, larger perturbations on the graph could lead to more informative triggers in a backdoor attack ({\em e.g.,} the pattern of the trigger is significant enough to influence the neural networks).
As we can see in most of the cases, when only one edge is perturbed, the attack success rate is quite low. As the perturbation size increases, the ASR can show a significant increase.
For example, on the Fingerprint dataset, the ASR is around 0.3, 0.25, 0.4, 0.45 for GCN, GIN, GSAGE and GAT when only one edge is perturbed. 
As the perturbation size increases to 5, the ASR also increases to 0.7, 0.75, 0.56 and 0.85 for GCN, GIN, GSAGE and GAT. 
Similar results can also be found on PROTEINS and WinMal.
Similar to the data poisoning rate, a larger edge perturbation number can not always guarantee higher attack effectiveness.
When the perturbation size increases from 5 to 7, the attack success rate does not increase much as shown on Fingerprint and PROTEINS. 
WinMal shows slightly increased ASR when the perturbation size increases. This is because WinMal has large graphs with the average edge number reaching to over 700. 
Hence, a larger perturbation could make the trigger become more informative. 
As for the clean accuracy drop shown in Fig. \ref{fig:trigger}, we can see that there is also no specific patterns and the overall CAD is less than 5\% in most of the cases.

\section{Randomized Subsampling Defense}

\begin{table}[ht]
    \centering
     \caption{Training with subsampling.}
    \begin{tabular}{|c|p{1.5cm}|p{1.5cm}|p{2cm}|}
    \hline
    Dataset & Clean \hspace{0.5cm} Accuracy & Backdoor Accuracy & Attack Success Rate \\\hline
         FRANKENSTEIN & 0.5121 &  0.4971  &  0.2775  \\ \hline 
         Fingerprint & 0.8116 & 0.8150 & 0.6379  \\  \hline 
         PROTEINS & 0.4081 & 0.4036  & 0.6125 \\ \hline 
         WinMal & 0.7766 & 0.7504 & 0.6557 \\ \hline 
        
    \end{tabular}
    \label{tab:subsampling_defense}
\end{table}

        

As the backdoor attacks on GNNs like TRAP and GTA are quite new  and their is no available mitigation specifically designed for graph domain. 
One possible way is to adopt the countermeasures from the other domains ({\em e.g.,} images) for defensing backdoor attacks against GNNs.
The existing defenses either perform inspections on the suspicious models \cite{,chen2019deepinspect,liu2019abs,wang2019neural} or detect the possibly poisoned inputs at inference time \cite{doan2020februus,gao2019strip,chen2018detecting,chou2018sentinet}.
For example, STRIP \cite{gao2019strip} performs input inspection via classification entropy difference on inputs with strong perturbation.
Februus \cite{doan2020februus} performs input purification where the trigger patterns in the images are detected and removed and before sending them to the neural networks.
Based on different principles, the countermeasures can some be classified as empirical  and certified defenses \cite{zhang2021backdoor}. 
Empirical defenses are designed for specific attacks and  they can be  bypassed when the attackers adopts adaptive attacks \cite{salem2020dynamic}. 
Certified defenses predict the inputs in a certain bound \cite{steinhardt2017certified, lecuyer2019certified}.

As for defense against GNN backdoor attacks, Xi et al. \cite{xi2021graph} extended the model inspection method NeuralCleanse (NC) \cite{wang2019neural} to defense GTA and adopted a minimum perturbation cost (MPC) measure for detection. They found that the MPC distributions of backdoor model attacked by GTA and benign models are similar, which shows the difficulty to defense GTA with MPC measures. 
Zhang et al. \cite{zhang2021backdoor} proposed to use randomized subsampling training as a certified defense  \cite{cao2017mitigating, liu2018towards,cohen2019certified}, which shows effectiveness in some cases. 
Therefore, we adopts the same subsampling  strategy to defense the attack of TRAP. 
We use randomized subsampling on the graph structure to defense the training datasets which may contain the perturbation trigger generated by TRAP.

Table \ref{tab:subsampling_defense} shows the results when training with subsampling for both clean  and backdoor GNN model. Here, we choose GCN model for a concrete since TRAP achieves the best  attack effectiveness on GCN.  
Following  Subgraph Backdoor \cite{zhang2021backdoor}, the subsampling ratio and sub-sampled graphs are set as 10\%  and 10, respectively. 
When trained without defense strategy, the clean accuracy (shown in Table \ref{tab:clean_model_acc}) is 0.6737, 0.8235,0.7195 and 0.8322 on the four datasets,  respectively.
However, when training with randomized subsampling for defense, the accuracy of both clean model and backdoor model drops significantly. 
For example, on FRANKENSTEIN and PROTEINS, the clean accuracy drops about 16\% and 31\%.
The attack success rate of backdoor GCN also drops which shows the impact of randomized subsampling.
However, the drop of ASR is not significant on most datasets except FRANKENSTEIN.
The reason is that our  trigger generated by TRAP only cause quite small  perturbation on the graph structure and therefore, the randomized subsampling can not complete destroy the trigger on the poisoned graphs.  
The results highlight that TRAP is still robust under the randomize subsampling defense strategy.

\section{Discussion}

Here, we discuss the potential reasons that lead to the effective and transferable attack of TRAP. 
As disclosed by previous works that adversarial samples show transferable property that some adversarial samples produced by one model could  mislead other models that are unseen and even their architectures are totally different \cite{papernot2016transferability,papernot2017practical}.
TRAP generates perturbation trigger for graph poisoning. The trigger-embedded graphs can be regarded as  adversarial examples except that the purpose of the perturbation trigger is for backdoor attack. 
Hence, one possible explanation for the transferable attack of TRAP is that  the vulnerability on a surrogate GNN model caused by the trigger  could transfer to other GNN models.

What is more, TRAP generates perturbation trigger which can impact the message passing of the surrogate GNN model  and lead its feature aggregation to the opposite: the  representation of a poisoned graph tend to be similar with the representation of a graph in the targeted attack class. 
And the different GNNs roughly have the similar type of message passing strategies in high-level aspect.  
As a result, the same perturbation trigger generated by a surrogate GCN model could also be transferred to the other GNNs  for backdoor attack.

\section{Conclusion}

In this paper, we  propose TRAP, a new backdoor attack against GNNs.
Compared with the existing backdoor attacks on GNNs, our attack excels in several aspects.
Our attack is more realistic, since we assume no control over the targeted GNN models. 
To be effective,   our attacks  are prepared on a surrogate model and  can be transferable to different GNN models through poisoning the training dataset. 
Moreover, our generated triggers are adaptive and without specific patterns for different graphs, and are dynamically generated via perturbing the graph structures.
We evaluate our attacks on real-world datasets, and the empirical studies demonstrate the effectiveness and evasiveness in backdoor attack against different GNN models.
Moreover, the transferability of our attacks is evaluated on four different GNNs, to which the attack is agnostic. 
TRAP is also tested on the  randomized subsampling based certified  defense strategy, the results show that the defense can not effectively defense  TRAP, which highlight the requirements for new defenses against backdoor attacks on GNNs.

\section{Acknowledgments}

This research was supported by Next Generation Technologies Fund programme with the Defence Science and Technology Group, Australia.



\bibliographystyle{ACM-Reference-Format}
\bibliography{bibfile.bib}


\begin{thebibliography}{54}


\ifx \showCODEN    \undefined \def \showCODEN     #1{\unskip}     \fi
\ifx \showDOI      \undefined \def \showDOI       #1{#1}\fi
\ifx \showISBNx    \undefined \def \showISBNx     #1{\unskip}     \fi
\ifx \showISBNxiii \undefined \def \showISBNxiii  #1{\unskip}     \fi
\ifx \showISSN     \undefined \def \showISSN      #1{\unskip}     \fi
\ifx \showLCCN     \undefined \def \showLCCN      #1{\unskip}     \fi
\ifx \shownote     \undefined \def \shownote      #1{#1}          \fi
\ifx \showarticletitle \undefined \def \showarticletitle #1{#1}   \fi
\ifx \showURL      \undefined \def \showURL       {\relax}        \fi
\providecommand\bibfield[2]{#2}
\providecommand\bibinfo[2]{#2}
\providecommand\natexlab[1]{#1}
\providecommand\showeprint[2][]{arXiv:#2}

\bibitem[\protect\citeauthoryear{Cao and Gong}{Cao and Gong}{2017}]%
        {cao2017mitigating}
\bibfield{author}{\bibinfo{person}{Xiaoyu Cao} {and}
  \bibinfo{person}{Neil~Zhenqiang Gong}.} \bibinfo{year}{2017}\natexlab{}.
\newblock \showarticletitle{Mitigating evasion attacks to deep neural networks
  via region-based classification}. In \bibinfo{booktitle}{\emph{Proceedings of
  the 33rd Annual Computer Security Applications Conference}}.
  \bibinfo{pages}{278--287}.
\newblock


\bibitem[\protect\citeauthoryear{Chakraborty, Alam, Dey, Chattopadhyay, and
  Mukhopadhyay}{Chakraborty et~al\mbox{.}}{2018}]%
        {chakraborty2018adversarial}
\bibfield{author}{\bibinfo{person}{Anirban Chakraborty},
  \bibinfo{person}{Manaar Alam}, \bibinfo{person}{Vishal Dey},
  \bibinfo{person}{Anupam Chattopadhyay}, {and} \bibinfo{person}{Debdeep
  Mukhopadhyay}.} \bibinfo{year}{2018}\natexlab{}.
\newblock \showarticletitle{Adversarial attacks and defences: A survey}.
\newblock \bibinfo{journal}{\emph{arXiv preprint arXiv:1810.00069}}
  (\bibinfo{year}{2018}).
\newblock


\bibitem[\protect\citeauthoryear{Chen, Carvalho, Baracaldo, Ludwig, Edwards,
  Lee, Molloy, and Srivastava}{Chen et~al\mbox{.}}{2018}]%
        {chen2018detecting}
\bibfield{author}{\bibinfo{person}{Bryant Chen}, \bibinfo{person}{Wilka
  Carvalho}, \bibinfo{person}{Nathalie Baracaldo}, \bibinfo{person}{Heiko
  Ludwig}, \bibinfo{person}{Benjamin Edwards}, \bibinfo{person}{Taesung Lee},
  \bibinfo{person}{Ian Molloy}, {and} \bibinfo{person}{Biplav Srivastava}.}
  \bibinfo{year}{2018}\natexlab{}.
\newblock \showarticletitle{Detecting backdoor attacks on deep neural networks
  by activation clustering}.
\newblock \bibinfo{journal}{\emph{arXiv preprint arXiv:1811.03728}}
  (\bibinfo{year}{2018}).
\newblock


\bibitem[\protect\citeauthoryear{Chen, Fu, Zhao, and Koushanfar}{Chen
  et~al\mbox{.}}{2019}]%
        {chen2019deepinspect}
\bibfield{author}{\bibinfo{person}{Huili Chen}, \bibinfo{person}{Cheng Fu},
  \bibinfo{person}{Jishen Zhao}, {and} \bibinfo{person}{Farinaz Koushanfar}.}
  \bibinfo{year}{2019}\natexlab{}.
\newblock \showarticletitle{DeepInspect: A Black-box Trojan Detection and
  Mitigation Framework for Deep Neural Networks.}. In
  \bibinfo{booktitle}{\emph{IJCAI}}, Vol.~\bibinfo{volume}{2}.
  \bibinfo{pages}{8}.
\newblock


\bibitem[\protect\citeauthoryear{Chou, Tram{\`e}r, Pellegrino, and Boneh}{Chou
  et~al\mbox{.}}{2018}]%
        {chou2018sentinet}
\bibfield{author}{\bibinfo{person}{Edward Chou}, \bibinfo{person}{Florian
  Tram{\`e}r}, \bibinfo{person}{Giancarlo Pellegrino}, {and}
  \bibinfo{person}{Dan Boneh}.} \bibinfo{year}{2018}\natexlab{}.
\newblock \showarticletitle{Sentinet: Detecting physical attacks against deep
  learning systems}.
\newblock  (\bibinfo{year}{2018}).
\newblock


\bibitem[\protect\citeauthoryear{Cohen, Rosenfeld, and Kolter}{Cohen
  et~al\mbox{.}}{2019}]%
        {cohen2019certified}
\bibfield{author}{\bibinfo{person}{Jeremy Cohen}, \bibinfo{person}{Elan
  Rosenfeld}, {and} \bibinfo{person}{Zico Kolter}.}
  \bibinfo{year}{2019}\natexlab{}.
\newblock \showarticletitle{Certified adversarial robustness via randomized
  smoothing}. In \bibinfo{booktitle}{\emph{International Conference on Machine
  Learning}}. PMLR, \bibinfo{pages}{1310--1320}.
\newblock


\bibitem[\protect\citeauthoryear{Dai, Li, Tian, Huang, Wang, Zhu, and Song}{Dai
  et~al\mbox{.}}{2018}]%
        {dai2018adversarial}
\bibfield{author}{\bibinfo{person}{Hanjun Dai}, \bibinfo{person}{Hui Li},
  \bibinfo{person}{Tian Tian}, \bibinfo{person}{Xin Huang},
  \bibinfo{person}{Lin Wang}, \bibinfo{person}{Jun Zhu}, {and}
  \bibinfo{person}{Le Song}.} \bibinfo{year}{2018}\natexlab{}.
\newblock \showarticletitle{Adversarial attack on graph structured data}. In
  \bibinfo{booktitle}{\emph{International conference on machine learning}}.
  PMLR, \bibinfo{pages}{1115--1124}.
\newblock


\bibitem[\protect\citeauthoryear{Doan, Abbasnejad, and Ranasinghe}{Doan
  et~al\mbox{.}}{2020}]%
        {doan2020februus}
\bibfield{author}{\bibinfo{person}{Bao~Gia Doan}, \bibinfo{person}{Ehsan
  Abbasnejad}, {and} \bibinfo{person}{Damith~C Ranasinghe}.}
  \bibinfo{year}{2020}\natexlab{}.
\newblock \showarticletitle{Februus: Input purification defense against trojan
  attacks on deep neural network systems}. In \bibinfo{booktitle}{\emph{Annual
  Computer Security Applications Conference}}. \bibinfo{pages}{897--912}.
\newblock


\bibitem[\protect\citeauthoryear{Doan, Xue, Ma, Abbasnejad, and
  Ranasinghe}{Doan et~al\mbox{.}}{2021}]%
        {bao2021tnt}
\bibfield{author}{\bibinfo{person}{Bao~Gia Doan}, \bibinfo{person}{Minhui Xue},
  \bibinfo{person}{Shiqing Ma}, \bibinfo{person}{Ehsan Abbasnejad}, {and}
  \bibinfo{person}{Damith~C. Ranasinghe}.} \bibinfo{year}{2021}\natexlab{}.
\newblock \showarticletitle{TnT {A}ttacks! {U}niversal Naturalistic Adversarial
  Patches Against Deep Neural Network Systems}.
\newblock  (\bibinfo{year}{2021}).
\newblock


\bibitem[\protect\citeauthoryear{Dobson and Doig}{Dobson and Doig}{2003}]%
        {dobson2003distinguishing}
\bibfield{author}{\bibinfo{person}{Paul~D Dobson} {and}
  \bibinfo{person}{Andrew~J Doig}.} \bibinfo{year}{2003}\natexlab{}.
\newblock \showarticletitle{Distinguishing enzyme structures from non-enzymes
  without alignments}.
\newblock \bibinfo{journal}{\emph{Journal of molecular biology}}
  \bibinfo{volume}{330}, \bibinfo{number}{4} (\bibinfo{year}{2003}),
  \bibinfo{pages}{771--783}.
\newblock


\bibitem[\protect\citeauthoryear{Gao, Xu, Wang, Chen, Ranasinghe, and
  Nepal}{Gao et~al\mbox{.}}{2019}]%
        {gao2019strip}
\bibfield{author}{\bibinfo{person}{Yansong Gao}, \bibinfo{person}{Change Xu},
  \bibinfo{person}{Derui Wang}, \bibinfo{person}{Shiping Chen},
  \bibinfo{person}{Damith~C Ranasinghe}, {and} \bibinfo{person}{Surya Nepal}.}
  \bibinfo{year}{2019}\natexlab{}.
\newblock \showarticletitle{Strip: A defence against trojan attacks on deep
  neural networks}. In \bibinfo{booktitle}{\emph{Proceedings of the 35th Annual
  Computer Security Applications Conference}}. \bibinfo{pages}{113--125}.
\newblock


\bibitem[\protect\citeauthoryear{Gilbert}{Gilbert}{1959}]%
        {gilbert1959random}
\bibfield{author}{\bibinfo{person}{Edgar~N Gilbert}.}
  \bibinfo{year}{1959}\natexlab{}.
\newblock \showarticletitle{Random graphs}.
\newblock \bibinfo{journal}{\emph{The Annals of Mathematical Statistics}}
  \bibinfo{volume}{30}, \bibinfo{number}{4} (\bibinfo{year}{1959}),
  \bibinfo{pages}{1141--1144}.
\newblock


\bibitem[\protect\citeauthoryear{Goodfellow, Shlens, and Szegedy}{Goodfellow
  et~al\mbox{.}}{2014}]%
        {goodfellow2014explaining}
\bibfield{author}{\bibinfo{person}{Ian~J Goodfellow}, \bibinfo{person}{Jonathon
  Shlens}, {and} \bibinfo{person}{Christian Szegedy}.}
  \bibinfo{year}{2014}\natexlab{}.
\newblock \showarticletitle{Explaining and harnessing adversarial examples}.
\newblock \bibinfo{journal}{\emph{arXiv preprint arXiv:1412.6572}}
  (\bibinfo{year}{2014}).
\newblock


\bibitem[\protect\citeauthoryear{Gu, Dolan-Gavitt, and Garg}{Gu
  et~al\mbox{.}}{2017}]%
        {gu2017badnets}
\bibfield{author}{\bibinfo{person}{Tianyu Gu}, \bibinfo{person}{Brendan
  Dolan-Gavitt}, {and} \bibinfo{person}{Siddharth Garg}.}
  \bibinfo{year}{2017}\natexlab{}.
\newblock \showarticletitle{Badnets: Identifying vulnerabilities in the machine
  learning model supply chain}.
\newblock \bibinfo{journal}{\emph{arXiv preprint arXiv:1708.06733}}
  (\bibinfo{year}{2017}).
\newblock


\bibitem[\protect\citeauthoryear{Gu, Liu, Dolan-Gavitt, and Garg}{Gu
  et~al\mbox{.}}{2019}]%
        {gu2019badnets}
\bibfield{author}{\bibinfo{person}{Tianyu Gu}, \bibinfo{person}{Kang Liu},
  \bibinfo{person}{Brendan Dolan-Gavitt}, {and} \bibinfo{person}{Siddharth
  Garg}.} \bibinfo{year}{2019}\natexlab{}.
\newblock \showarticletitle{Badnets: Evaluating backdooring attacks on deep
  neural networks}.
\newblock \bibinfo{journal}{\emph{IEEE Access}}  \bibinfo{volume}{7}
  (\bibinfo{year}{2019}), \bibinfo{pages}{47230--47244}.
\newblock


\bibitem[\protect\citeauthoryear{Hamilton, Ying, and Leskovec}{Hamilton
  et~al\mbox{.}}{2017}]%
        {hamilton2017inductive}
\bibfield{author}{\bibinfo{person}{Will Hamilton}, \bibinfo{person}{Zhitao
  Ying}, {and} \bibinfo{person}{Jure Leskovec}.}
  \bibinfo{year}{2017}\natexlab{}.
\newblock \showarticletitle{Inductive representation learning on large graphs}.
\newblock \bibinfo{journal}{\emph{Advances in neural information processing
  systems}}  \bibinfo{volume}{30} (\bibinfo{year}{2017}).
\newblock


\bibitem[\protect\citeauthoryear{Jin, Li, Xu, Wang, Ji, Aggarwal, and Tang}{Jin
  et~al\mbox{.}}{2021}]%
        {jin2020graph}
\bibfield{author}{\bibinfo{person}{Wei Jin}, \bibinfo{person}{Yaxing Li},
  \bibinfo{person}{Han Xu}, \bibinfo{person}{Yiqi Wang},
  \bibinfo{person}{Shuiwang Ji}, \bibinfo{person}{Charu Aggarwal}, {and}
  \bibinfo{person}{Jiliang Tang}.} \bibinfo{year}{2021}\natexlab{}.
\newblock \showarticletitle{Adversarial Attacks and Defenses on Graphs}.
\newblock \bibinfo{journal}{\emph{SIGKDD Explor. Newsl.}} \bibinfo{volume}{22},
  \bibinfo{number}{2} (\bibinfo{date}{jan} \bibinfo{year}{2021}),
  \bibinfo{pages}{19–34}.
\newblock
\showISSN{1931-0145}
\urldef\tempurl%
\url{https://doi.org/10.1145/3447556.3447566}
\showDOI{\tempurl}


\bibitem[\protect\citeauthoryear{Kipf and Welling}{Kipf and Welling}{2016}]%
        {kipf2016semi}
\bibfield{author}{\bibinfo{person}{Thomas~N Kipf} {and} \bibinfo{person}{Max
  Welling}.} \bibinfo{year}{2016}\natexlab{}.
\newblock \showarticletitle{Semi-supervised classification with graph
  convolutional networks}.
\newblock \bibinfo{journal}{\emph{arXiv preprint arXiv:1609.02907}}
  (\bibinfo{year}{2016}).
\newblock


\bibitem[\protect\citeauthoryear{Lecuyer, Atlidakis, Geambasu, Hsu, and
  Jana}{Lecuyer et~al\mbox{.}}{2019}]%
        {lecuyer2019certified}
\bibfield{author}{\bibinfo{person}{Mathias Lecuyer}, \bibinfo{person}{Vaggelis
  Atlidakis}, \bibinfo{person}{Roxana Geambasu}, \bibinfo{person}{Daniel Hsu},
  {and} \bibinfo{person}{Suman Jana}.} \bibinfo{year}{2019}\natexlab{}.
\newblock \showarticletitle{Certified robustness to adversarial examples with
  differential privacy}. In \bibinfo{booktitle}{\emph{2019 IEEE Symposium on
  Security and Privacy (SP)}}. IEEE, \bibinfo{pages}{656--672}.
\newblock


\bibitem[\protect\citeauthoryear{Li, Li, Wu, Li, He, and Lyu}{Li
  et~al\mbox{.}}{2021}]%
        {li2021invisible}
\bibfield{author}{\bibinfo{person}{Yuezun Li}, \bibinfo{person}{Yiming Li},
  \bibinfo{person}{Baoyuan Wu}, \bibinfo{person}{Longkang Li},
  \bibinfo{person}{Ran He}, {and} \bibinfo{person}{Siwei Lyu}.}
  \bibinfo{year}{2021}\natexlab{}.
\newblock \showarticletitle{Invisible backdoor attack with sample-specific
  triggers}. In \bibinfo{booktitle}{\emph{Proceedings of the IEEE/CVF
  International Conference on Computer Vision}}. \bibinfo{pages}{16463--16472}.
\newblock


\bibitem[\protect\citeauthoryear{Liao, Zhong, Squicciarini, Zhu, and
  Miller}{Liao et~al\mbox{.}}{2018}]%
        {liao2018backdoor}
\bibfield{author}{\bibinfo{person}{Cong Liao}, \bibinfo{person}{Haoti Zhong},
  \bibinfo{person}{Anna Squicciarini}, \bibinfo{person}{Sencun Zhu}, {and}
  \bibinfo{person}{David Miller}.} \bibinfo{year}{2018}\natexlab{}.
\newblock \showarticletitle{Backdoor embedding in convolutional neural network
  models via invisible perturbation}.
\newblock \bibinfo{journal}{\emph{arXiv preprint arXiv:1808.10307}}
  (\bibinfo{year}{2018}).
\newblock


\bibitem[\protect\citeauthoryear{Liu, Cheng, Zhang, and Hsieh}{Liu
  et~al\mbox{.}}{2018}]%
        {liu2018towards}
\bibfield{author}{\bibinfo{person}{Xuanqing Liu}, \bibinfo{person}{Minhao
  Cheng}, \bibinfo{person}{Huan Zhang}, {and} \bibinfo{person}{Cho-Jui Hsieh}.}
  \bibinfo{year}{2018}\natexlab{}.
\newblock \showarticletitle{Towards robust neural networks via random
  self-ensemble}. In \bibinfo{booktitle}{\emph{Proceedings of the European
  Conference on Computer Vision (ECCV)}}. \bibinfo{pages}{369--385}.
\newblock


\bibitem[\protect\citeauthoryear{Liu, Chen, Liu, and Song}{Liu
  et~al\mbox{.}}{2016}]%
        {liu2016delving}
\bibfield{author}{\bibinfo{person}{Yanpei Liu}, \bibinfo{person}{Xinyun Chen},
  \bibinfo{person}{Chang Liu}, {and} \bibinfo{person}{Dawn Song}.}
  \bibinfo{year}{2016}\natexlab{}.
\newblock \showarticletitle{Delving into transferable adversarial examples and
  black-box attacks}.
\newblock \bibinfo{journal}{\emph{arXiv preprint arXiv:1611.02770}}
  (\bibinfo{year}{2016}).
\newblock


\bibitem[\protect\citeauthoryear{Liu, Lee, Tao, Ma, Aafer, and Zhang}{Liu
  et~al\mbox{.}}{2019}]%
        {liu2019abs}
\bibfield{author}{\bibinfo{person}{Yingqi Liu}, \bibinfo{person}{Wen-Chuan
  Lee}, \bibinfo{person}{Guanhong Tao}, \bibinfo{person}{Shiqing Ma},
  \bibinfo{person}{Yousra Aafer}, {and} \bibinfo{person}{Xiangyu Zhang}.}
  \bibinfo{year}{2019}\natexlab{}.
\newblock \showarticletitle{Abs: Scanning neural networks for back-doors by
  artificial brain stimulation}. In \bibinfo{booktitle}{\emph{Proceedings of
  the 2019 ACM SIGSAC Conference on Computer and Communications Security}}.
  \bibinfo{pages}{1265--1282}.
\newblock


\bibitem[\protect\citeauthoryear{Liu, Xie, and Srivastava}{Liu
  et~al\mbox{.}}{2017}]%
        {liu2017neural}
\bibfield{author}{\bibinfo{person}{Yuntao Liu}, \bibinfo{person}{Yang Xie},
  {and} \bibinfo{person}{Ankur Srivastava}.} \bibinfo{year}{2017}\natexlab{}.
\newblock \showarticletitle{Neural trojans}. In \bibinfo{booktitle}{\emph{2017
  IEEE International Conference on Computer Design (ICCD)}}. IEEE,
  \bibinfo{pages}{45--48}.
\newblock


\bibitem[\protect\citeauthoryear{Madry, Makelov, Schmidt, Tsipras, and
  Vladu}{Madry et~al\mbox{.}}{2017}]%
        {madry2017towards}
\bibfield{author}{\bibinfo{person}{Aleksander Madry},
  \bibinfo{person}{Aleksandar Makelov}, \bibinfo{person}{Ludwig Schmidt},
  \bibinfo{person}{Dimitris Tsipras}, {and} \bibinfo{person}{Adrian Vladu}.}
  \bibinfo{year}{2017}\natexlab{}.
\newblock \showarticletitle{Towards deep learning models resistant to
  adversarial attacks}.
\newblock \bibinfo{journal}{\emph{arXiv preprint arXiv:1706.06083}}
  (\bibinfo{year}{2017}).
\newblock


\bibitem[\protect\citeauthoryear{Moosavi-Dezfooli, Fawzi, and
  Frossard}{Moosavi-Dezfooli et~al\mbox{.}}{2016}]%
        {moosavi2016deepfool}
\bibfield{author}{\bibinfo{person}{Seyed-Mohsen Moosavi-Dezfooli},
  \bibinfo{person}{Alhussein Fawzi}, {and} \bibinfo{person}{Pascal Frossard}.}
  \bibinfo{year}{2016}\natexlab{}.
\newblock \showarticletitle{Deepfool: a simple and accurate method to fool deep
  neural networks}. In \bibinfo{booktitle}{\emph{Proceedings of the IEEE
  conference on computer vision and pattern recognition}}.
  \bibinfo{pages}{2574--2582}.
\newblock


\bibitem[\protect\citeauthoryear{Morris, Kriege, Bause, Kersting, Mutzel, and
  Neumann}{Morris et~al\mbox{.}}{2020}]%
        {morris2020tudataset}
\bibfield{author}{\bibinfo{person}{Christopher Morris}, \bibinfo{person}{Nils~M
  Kriege}, \bibinfo{person}{Franka Bause}, \bibinfo{person}{Kristian Kersting},
  \bibinfo{person}{Petra Mutzel}, {and} \bibinfo{person}{Marion Neumann}.}
  \bibinfo{year}{2020}\natexlab{}.
\newblock \showarticletitle{Tudataset: A collection of benchmark datasets for
  learning with graphs}.
\newblock \bibinfo{journal}{\emph{arXiv preprint arXiv:2007.08663}}
  (\bibinfo{year}{2020}).
\newblock


\bibitem[\protect\citeauthoryear{Mu{\~n}oz-Gonz{\'a}lez, Pfitzner, Russo,
  Carnerero-Cano, and Lupu}{Mu{\~n}oz-Gonz{\'a}lez et~al\mbox{.}}{2019}]%
        {munoz2019poisoning}
\bibfield{author}{\bibinfo{person}{Luis Mu{\~n}oz-Gonz{\'a}lez},
  \bibinfo{person}{Bjarne Pfitzner}, \bibinfo{person}{Matteo Russo},
  \bibinfo{person}{Javier Carnerero-Cano}, {and} \bibinfo{person}{Emil~C
  Lupu}.} \bibinfo{year}{2019}\natexlab{}.
\newblock \showarticletitle{Poisoning attacks with generative adversarial
  nets}.
\newblock \bibinfo{journal}{\emph{arXiv preprint arXiv:1906.07773}}
  (\bibinfo{year}{2019}).
\newblock


\bibitem[\protect\citeauthoryear{Orsini, Frasconi, and De~Raedt}{Orsini
  et~al\mbox{.}}{2015}]%
        {orsini2015graph}
\bibfield{author}{\bibinfo{person}{Francesco Orsini}, \bibinfo{person}{Paolo
  Frasconi}, {and} \bibinfo{person}{Luc De~Raedt}.}
  \bibinfo{year}{2015}\natexlab{}.
\newblock \showarticletitle{Graph invariant kernels}. In
  \bibinfo{booktitle}{\emph{Twenty-Fourth International Joint Conference on
  Artificial Intelligence}}.
\newblock


\bibitem[\protect\citeauthoryear{Papernot, McDaniel, and Goodfellow}{Papernot
  et~al\mbox{.}}{2016a}]%
        {papernot2016transferability}
\bibfield{author}{\bibinfo{person}{Nicolas Papernot}, \bibinfo{person}{Patrick
  McDaniel}, {and} \bibinfo{person}{Ian Goodfellow}.}
  \bibinfo{year}{2016}\natexlab{a}.
\newblock \showarticletitle{Transferability in machine learning: from phenomena
  to black-box attacks using adversarial samples}.
\newblock \bibinfo{journal}{\emph{arXiv preprint arXiv:1605.07277}}
  (\bibinfo{year}{2016}).
\newblock


\bibitem[\protect\citeauthoryear{Papernot, McDaniel, Goodfellow, Jha, Celik,
  and Swami}{Papernot et~al\mbox{.}}{2017}]%
        {papernot2017practical}
\bibfield{author}{\bibinfo{person}{Nicolas Papernot}, \bibinfo{person}{Patrick
  McDaniel}, \bibinfo{person}{Ian Goodfellow}, \bibinfo{person}{Somesh Jha},
  \bibinfo{person}{Z~Berkay Celik}, {and} \bibinfo{person}{Ananthram Swami}.}
  \bibinfo{year}{2017}\natexlab{}.
\newblock \showarticletitle{Practical black-box attacks against machine
  learning}. In \bibinfo{booktitle}{\emph{Proceedings of the 2017 ACM on Asia
  conference on computer and communications security}}.
  \bibinfo{pages}{506--519}.
\newblock


\bibitem[\protect\citeauthoryear{Papernot, McDaniel, Jha, Fredrikson, Celik,
  and Swami}{Papernot et~al\mbox{.}}{2016b}]%
        {papernot2016limitations}
\bibfield{author}{\bibinfo{person}{Nicolas Papernot}, \bibinfo{person}{Patrick
  McDaniel}, \bibinfo{person}{Somesh Jha}, \bibinfo{person}{Matt Fredrikson},
  \bibinfo{person}{Z~Berkay Celik}, {and} \bibinfo{person}{Ananthram Swami}.}
  \bibinfo{year}{2016}\natexlab{b}.
\newblock \showarticletitle{The limitations of deep learning in adversarial
  settings}. In \bibinfo{booktitle}{\emph{2016 IEEE European symposium on
  security and privacy (EuroS\&P)}}. IEEE, \bibinfo{pages}{372--387}.
\newblock


\bibitem[\protect\citeauthoryear{Ranveer and Hiray}{Ranveer and Hiray}{2015}]%
        {ranveer2015comparative}
\bibfield{author}{\bibinfo{person}{Smita Ranveer} {and}
  \bibinfo{person}{Swapnaja Hiray}.} \bibinfo{year}{2015}\natexlab{}.
\newblock \showarticletitle{Comparative analysis of feature extraction methods
  of malware detection}.
\newblock \bibinfo{journal}{\emph{International Journal of Computer
  Applications}} \bibinfo{volume}{120}, \bibinfo{number}{5}
  (\bibinfo{year}{2015}).
\newblock


\bibitem[\protect\citeauthoryear{Salem, Wen, Backes, Ma, and Zhang}{Salem
  et~al\mbox{.}}{2020}]%
        {salem2020dynamic}
\bibfield{author}{\bibinfo{person}{Ahmed Salem}, \bibinfo{person}{Rui Wen},
  \bibinfo{person}{Michael Backes}, \bibinfo{person}{Shiqing Ma}, {and}
  \bibinfo{person}{Yang Zhang}.} \bibinfo{year}{2020}\natexlab{}.
\newblock \showarticletitle{Dynamic backdoor attacks against machine learning
  models}.
\newblock \bibinfo{journal}{\emph{arXiv preprint arXiv:2003.03675}}
  (\bibinfo{year}{2020}).
\newblock


\bibitem[\protect\citeauthoryear{Steinhardt, Koh, and Liang}{Steinhardt
  et~al\mbox{.}}{2017}]%
        {steinhardt2017certified}
\bibfield{author}{\bibinfo{person}{Jacob Steinhardt}, \bibinfo{person}{Pang
  Wei~W Koh}, {and} \bibinfo{person}{Percy~S Liang}.}
  \bibinfo{year}{2017}\natexlab{}.
\newblock \showarticletitle{Certified defenses for data poisoning attacks}.
\newblock \bibinfo{journal}{\emph{Advances in neural information processing
  systems}}  \bibinfo{volume}{30} (\bibinfo{year}{2017}).
\newblock


\bibitem[\protect\citeauthoryear{Sun, Dou, Yang, Wang, Yu, He, and Li}{Sun
  et~al\mbox{.}}{2018}]%
        {sun2018adversarial}
\bibfield{author}{\bibinfo{person}{Lichao Sun}, \bibinfo{person}{Yingtong Dou},
  \bibinfo{person}{Carl Yang}, \bibinfo{person}{Ji Wang},
  \bibinfo{person}{Philip~S Yu}, \bibinfo{person}{Lifang He}, {and}
  \bibinfo{person}{Bo Li}.} \bibinfo{year}{2018}\natexlab{}.
\newblock \showarticletitle{Adversarial attack and defense on graph data: A
  survey}.
\newblock \bibinfo{journal}{\emph{arXiv preprint arXiv:1812.10528}}
  (\bibinfo{year}{2018}).
\newblock


\bibitem[\protect\citeauthoryear{Veli{\v{c}}kovi{\'c}, Cucurull, Casanova,
  Romero, Lio, and Bengio}{Veli{\v{c}}kovi{\'c} et~al\mbox{.}}{2017}]%
        {velivckovic2017graph}
\bibfield{author}{\bibinfo{person}{Petar Veli{\v{c}}kovi{\'c}},
  \bibinfo{person}{Guillem Cucurull}, \bibinfo{person}{Arantxa Casanova},
  \bibinfo{person}{Adriana Romero}, \bibinfo{person}{Pietro Lio}, {and}
  \bibinfo{person}{Yoshua Bengio}.} \bibinfo{year}{2017}\natexlab{}.
\newblock \showarticletitle{Graph attention networks}.
\newblock \bibinfo{journal}{\emph{arXiv preprint arXiv:1710.10903}}
  (\bibinfo{year}{2017}).
\newblock


\bibitem[\protect\citeauthoryear{Wan, Kenlay, Ru, Blaas, Osborne, and Dong}{Wan
  et~al\mbox{.}}{2021}]%
        {wan2021attacking}
\bibfield{author}{\bibinfo{person}{Xingchen Wan}, \bibinfo{person}{Henry
  Kenlay}, \bibinfo{person}{Binxin Ru}, \bibinfo{person}{Arno Blaas},
  \bibinfo{person}{Michael Osborne}, {and} \bibinfo{person}{Xiaowen Dong}.}
  \bibinfo{year}{2021}\natexlab{}.
\newblock \showarticletitle{Attacking Graph Classification via Bayesian
  Optimisation}. In \bibinfo{booktitle}{\emph{ICML 2021 Workshop on Adversarial
  Machine Learning}}.
\newblock


\bibitem[\protect\citeauthoryear{Wang and Gong}{Wang and Gong}{2019}]%
        {wang2019attacking}
\bibfield{author}{\bibinfo{person}{Binghui Wang} {and}
  \bibinfo{person}{Neil~Zhenqiang Gong}.} \bibinfo{year}{2019}\natexlab{}.
\newblock \showarticletitle{Attacking graph-based classification via
  manipulating the graph structure}. In \bibinfo{booktitle}{\emph{Proceedings
  of the 2019 ACM SIGSAC Conference on Computer and Communications Security}}.
  \bibinfo{pages}{2023--2040}.
\newblock


\bibitem[\protect\citeauthoryear{Wang, Yao, Shan, Li, Viswanath, Zheng, and
  Zhao}{Wang et~al\mbox{.}}{2019}]%
        {wang2019neural}
\bibfield{author}{\bibinfo{person}{Bolun Wang}, \bibinfo{person}{Yuanshun Yao},
  \bibinfo{person}{Shawn Shan}, \bibinfo{person}{Huiying Li},
  \bibinfo{person}{Bimal Viswanath}, \bibinfo{person}{Haitao Zheng}, {and}
  \bibinfo{person}{Ben~Y Zhao}.} \bibinfo{year}{2019}\natexlab{}.
\newblock \showarticletitle{Neural cleanse: Identifying and mitigating backdoor
  attacks in neural networks}. In \bibinfo{booktitle}{\emph{2019 IEEE Symposium
  on Security and Privacy (SP)}}. IEEE, \bibinfo{pages}{707--723}.
\newblock


\bibitem[\protect\citeauthoryear{Watson and Wilson}{Watson and Wilson}{1992}]%
        {watson1992nist}
\bibfield{author}{\bibinfo{person}{Craig~I Watson} {and}
  \bibinfo{person}{Charles~L Wilson}.} \bibinfo{year}{1992}\natexlab{}.
\newblock \showarticletitle{NIST special database 4}.
\newblock \bibinfo{journal}{\emph{Fingerprint Database, National Institute of
  Standards and Technology}} \bibinfo{volume}{17}, \bibinfo{number}{77}
  (\bibinfo{year}{1992}), \bibinfo{pages}{5}.
\newblock


\bibitem[\protect\citeauthoryear{Wu, Wang, Tyshetskiy, Docherty, Lu, and
  Zhu}{Wu et~al\mbox{.}}{2019}]%
        {wu2019adversarial}
\bibfield{author}{\bibinfo{person}{Huijun Wu}, \bibinfo{person}{Chen Wang},
  \bibinfo{person}{Yuriy Tyshetskiy}, \bibinfo{person}{Andrew Docherty},
  \bibinfo{person}{Kai Lu}, {and} \bibinfo{person}{Liming Zhu}.}
  \bibinfo{year}{2019}\natexlab{}.
\newblock \showarticletitle{Adversarial examples on graph data: Deep insights
  into attack and defense}.
\newblock \bibinfo{journal}{\emph{arXiv preprint arXiv:1903.01610}}
  (\bibinfo{year}{2019}).
\newblock


\bibitem[\protect\citeauthoryear{Xi, Pang, Ji, and Wang}{Xi
  et~al\mbox{.}}{2021}]%
        {xi2021graph}
\bibfield{author}{\bibinfo{person}{Zhaohan Xi}, \bibinfo{person}{Ren Pang},
  \bibinfo{person}{Shouling Ji}, {and} \bibinfo{person}{Ting Wang}.}
  \bibinfo{year}{2021}\natexlab{}.
\newblock \showarticletitle{Graph backdoor}. In \bibinfo{booktitle}{\emph{30th
  $\{$USENIX$\}$ Security Symposium ($\{$USENIX$\}$ Security 21)}}.
\newblock


\bibitem[\protect\citeauthoryear{Xu, Chen, Liu, Chen, Weng, Hong, and Lin}{Xu
  et~al\mbox{.}}{2019}]%
        {xu2019topology}
\bibfield{author}{\bibinfo{person}{Kaidi Xu}, \bibinfo{person}{Hongge Chen},
  \bibinfo{person}{Sijia Liu}, \bibinfo{person}{Pin-Yu Chen},
  \bibinfo{person}{Tsui-Wei Weng}, \bibinfo{person}{Mingyi Hong}, {and}
  \bibinfo{person}{Xue Lin}.} \bibinfo{year}{2019}\natexlab{}.
\newblock \showarticletitle{Topology attack and defense for graph neural
  networks: An optimization perspective}.
\newblock \bibinfo{journal}{\emph{arXiv preprint arXiv:1906.04214}}
  (\bibinfo{year}{2019}).
\newblock


\bibitem[\protect\citeauthoryear{Xu, Hu, Leskovec, and Jegelka}{Xu
  et~al\mbox{.}}{2018}]%
        {xu2018powerful}
\bibfield{author}{\bibinfo{person}{Keyulu Xu}, \bibinfo{person}{Weihua Hu},
  \bibinfo{person}{Jure Leskovec}, {and} \bibinfo{person}{Stefanie Jegelka}.}
  \bibinfo{year}{2018}\natexlab{}.
\newblock \showarticletitle{How powerful are graph neural networks?}
\newblock \bibinfo{journal}{\emph{arXiv preprint arXiv:1810.00826}}
  (\bibinfo{year}{2018}).
\newblock


\bibitem[\protect\citeauthoryear{Yang, Verma, Cai, Jiang, Yu, Chen, and
  Yu}{Yang et~al\mbox{.}}{2021}]%
        {yang2021variational}
\bibfield{author}{\bibinfo{person}{Shuiqiao Yang}, \bibinfo{person}{Sunny
  Verma}, \bibinfo{person}{Borui Cai}, \bibinfo{person}{Jiaojiao Jiang},
  \bibinfo{person}{Kun Yu}, \bibinfo{person}{Fang Chen}, {and}
  \bibinfo{person}{Shui Yu}.} \bibinfo{year}{2021}\natexlab{}.
\newblock \showarticletitle{Variational Co-embedding Learning for Attributed
  Network Clustering}.
\newblock \bibinfo{journal}{\emph{arXiv preprint arXiv:2104.07295}}
  (\bibinfo{year}{2021}).
\newblock


\bibitem[\protect\citeauthoryear{Yuan, Zhang, Jia, Tan, Xue, and Shan}{Yuan
  et~al\mbox{.}}{2021}]%
        {yuan2021meta}
\bibfield{author}{\bibinfo{person}{Zheng Yuan}, \bibinfo{person}{Jie Zhang},
  \bibinfo{person}{Yunpei Jia}, \bibinfo{person}{Chuanqi Tan},
  \bibinfo{person}{Tao Xue}, {and} \bibinfo{person}{Shiguang Shan}.}
  \bibinfo{year}{2021}\natexlab{}.
\newblock \showarticletitle{Meta gradient adversarial attack}. In
  \bibinfo{booktitle}{\emph{Proceedings of the IEEE/CVF International
  Conference on Computer Vision}}. \bibinfo{pages}{7748--7757}.
\newblock


\bibitem[\protect\citeauthoryear{Zang, Xie, Chen, and Yuan}{Zang
  et~al\mbox{.}}{2020}]%
        {zang2020graph}
\bibfield{author}{\bibinfo{person}{Xiao Zang}, \bibinfo{person}{Yi Xie},
  \bibinfo{person}{Jie Chen}, {and} \bibinfo{person}{Bo Yuan}.}
  \bibinfo{year}{2020}\natexlab{}.
\newblock \showarticletitle{Graph universal adversarial attacks: A few bad
  actors ruin graph learning models}.
\newblock \bibinfo{journal}{\emph{arXiv preprint arXiv:2002.04784}}
  (\bibinfo{year}{2020}).
\newblock


\bibitem[\protect\citeauthoryear{Zhang, Jia, Wang, and Gong}{Zhang
  et~al\mbox{.}}{2021}]%
        {zhang2021backdoor}
\bibfield{author}{\bibinfo{person}{Zaixi Zhang}, \bibinfo{person}{Jinyuan Jia},
  \bibinfo{person}{Binghui Wang}, {and} \bibinfo{person}{Neil~Zhenqiang Gong}.}
  \bibinfo{year}{2021}\natexlab{}.
\newblock \showarticletitle{Backdoor attacks to graph neural networks}. In
  \bibinfo{booktitle}{\emph{Proceedings of the 26th ACM Symposium on Access
  Control Models and Technologies}}. \bibinfo{pages}{15--26}.
\newblock


\bibitem[\protect\citeauthoryear{Zhao, Ma, Zheng, Bailey, Chen, and Jiang}{Zhao
  et~al\mbox{.}}{2020}]%
        {zhao2020clean}
\bibfield{author}{\bibinfo{person}{Shihao Zhao}, \bibinfo{person}{Xingjun Ma},
  \bibinfo{person}{Xiang Zheng}, \bibinfo{person}{James Bailey},
  \bibinfo{person}{Jingjing Chen}, {and} \bibinfo{person}{Yu-Gang Jiang}.}
  \bibinfo{year}{2020}\natexlab{}.
\newblock \showarticletitle{Clean-label backdoor attacks on video recognition
  models}. In \bibinfo{booktitle}{\emph{Proceedings of the IEEE/CVF Conference
  on Computer Vision and Pattern Recognition}}. \bibinfo{pages}{14443--14452}.
\newblock


\bibitem[\protect\citeauthoryear{Zhou, Cui, Hu, Zhang, Yang, Liu, Wang, Li, and
  Sun}{Zhou et~al\mbox{.}}{2020}]%
        {zhou2020graph}
\bibfield{author}{\bibinfo{person}{Jie Zhou}, \bibinfo{person}{Ganqu Cui},
  \bibinfo{person}{Shengding Hu}, \bibinfo{person}{Zhengyan Zhang},
  \bibinfo{person}{Cheng Yang}, \bibinfo{person}{Zhiyuan Liu},
  \bibinfo{person}{Lifeng Wang}, \bibinfo{person}{Changcheng Li}, {and}
  \bibinfo{person}{Maosong Sun}.} \bibinfo{year}{2020}\natexlab{}.
\newblock \showarticletitle{Graph neural networks: A review of methods and
  applications}.
\newblock \bibinfo{journal}{\emph{AI Open}}  \bibinfo{volume}{1}
  (\bibinfo{year}{2020}), \bibinfo{pages}{57--81}.
\newblock


\bibitem[\protect\citeauthoryear{Z{\"u}gner, Akbarnejad, and
  G{\"u}nnemann}{Z{\"u}gner et~al\mbox{.}}{2018}]%
        {zugner2018adversarial}
\bibfield{author}{\bibinfo{person}{Daniel Z{\"u}gner}, \bibinfo{person}{Amir
  Akbarnejad}, {and} \bibinfo{person}{Stephan G{\"u}nnemann}.}
  \bibinfo{year}{2018}\natexlab{}.
\newblock \showarticletitle{Adversarial attacks on neural networks for graph
  data}. In \bibinfo{booktitle}{\emph{Proceedings of the 24th ACM SIGKDD
  International Conference on Knowledge Discovery \& Data Mining}}.
  \bibinfo{pages}{2847--2856}.
\newblock


\bibitem[\protect\citeauthoryear{Z{\"u}gner and G{\"u}nnemann}{Z{\"u}gner and
  G{\"u}nnemann}{2019}]%
        {zugner2019adversarial}
\bibfield{author}{\bibinfo{person}{Daniel Z{\"u}gner} {and}
  \bibinfo{person}{Stephan G{\"u}nnemann}.} \bibinfo{year}{2019}\natexlab{}.
\newblock \showarticletitle{Adversarial attacks on graph neural networks via
  meta learning}.
\newblock \bibinfo{journal}{\emph{arXiv preprint arXiv:1902.08412}}
  (\bibinfo{year}{2019}).
\newblock


\end{thebibliography}
\end{document}